\documentclass[preprint,twocolumn]{aastex701}

\usepackage{graphicx}

\usepackage{amsmath,amssymb}
\usepackage{txfonts}
\usepackage{natbib}
\usepackage{subfigure}

\usepackage[normalem]{ulem}

\usepackage{xcolor}  

\usepackage{hyperref}
\begin{document}

\title{A Self-Consistent Model of Kinetic Alfv\'{e}n Solitons in Pulsar Wind Plasma: Linking Soliton Characteristics to Pulsar Observables}

\author[0000-0003-0289-2818]{Manpreet Singh}
\affiliation{School of Computing and Artificial Intelligence, Southwest Jiaotong University, Chengdu-610031, PR China.}
\affiliation{School of Physical Science and Technology, Southwest Jiaotong University, Chengdu-610031, PR China}
\email[show]{singhmanpreet185@gmail.com}  

\author[0000-0002-7421-1324]{Geetika Slathia} 
\affiliation{Department of Physics, Guru Nanak Dev University, Amritsar-143005, India}
\affiliation{Department of Physics, Government College for Women, Udhampur-182101, Jammu and Kashmir, India}
\email{gslathia93@gmail.com}

\author[0000-0001-7739-8377]{N. S. Saini}
\affiliation{Department of Physics, Guru Nanak Dev University, Amritsar-143005, India}
\email{nssaini@yahoo.com}

\author[0000-0003-1039-9521]{Siming Liu}
\affiliation{School of Physical Science and Technology, Southwest Jiaotong University, Chengdu-610031, PR China}
\email[]{liusm@swjtu.edu.cn}

\footnotetext[1]{Corresponding author: Manpreet Singh}

\shorttitle{KA Solitons in Pulsar Wind Plasma}
\shortauthors{Singh et al.}

\begin{abstract}

The nonlinear 
\end{abstract}


\begin{abstract}
A self-consistent model is presented for the formation and propagation of kinetic Alfv\'en (KA) solitons in mass-loaded filaments within the pulsar wind, where a magnetized electron--positron--ion plasma flows along open magnetic field lines beyond the light cylinder. Using a reductive perturbation approach, we derive a Korteweg--de Vries (KdV) equation governing the nonlinear evolution of KA solitons in this environment. The soliton amplitude and width depend sensitively on key pulsar observables, including spin period, spin-down rate, and pair multiplicity, as well as on plasma composition and suprathermal particle distributions. Heavy ion species such as Fe$^{26+}$ produce significantly broader solitons through enhanced inertia and dispersion, while increasing pair multiplicity leads to smaller solitons through stronger screening. More oblique propagation (larger $\theta$) yields wider but lower-amplitude solitons, whereas more thermalized pair plasmas (higher $\kappa$) support taller and broader structures.
A population-level analysis of 1174 pulsars quantifies the physical scales of these nonlinear structures, showing that millisecond pulsars host the most compact solitons, whereas slower pulsars support broader structures. 
Within the adopted admissible finite-\(\beta\) regime, this work links soliton properties to measurable pulsar parameters and provides a self-consistent framework for characterizing localized nonlinear plasma structures in finite-magnetization regions of pulsar winds and for assessing their role in modulating the local plasma environment. 

\end{abstract}

\section{Introduction}
\label{Introduction}

Pulsars are rapidly rotating neutron stars with typical masses of $\sim1.4\,M_\odot$ (where $M_\odot$ denotes the solar mass), radii of $\sim$10$-$15 km, and rotation periods ranging from milliseconds to seconds \citep{Michel91}. 
Their surface magnetic fields (often $\sim10^{12}\,$G), combined with fast spin, generate enormous induced electric fields that may exceed $10^{12}$ V/m \citep{GOLDREICHPulsar1969, Benacek2024} near the star surface. 
Consequently, the plasma near the pulsar consists of a complex mixture of electrons, positrons, and a minor component of ions extracted from the neutron star’s surface, forming what is often termed as electron–positron–ion (e--p--i) plasma.
The foundational work of Goldreich and Julian \citep{GOLDREICHPulsar1969} established that the rotation of a strongly magnetized neutron star induces a charge-separated magnetosphere with a characteristic density, known as the Goldreich–Julian density $n_\mathrm{GJ}$, which serves as a baseline for plasma density. 
Subsequent studies \citep{Ruderman75, Arons83} have demonstrated that pair cascades near the polar caps can enhance the density of electrons and positrons by several orders of magnitude ($\sim 1-10^6$;  \citet{DEJAGERGammaRay1996,DEJAGERLower2007,KISAKATeV2012a}) relative to this baseline density ($n_\mathrm{GJ}$), while the ion density remains close to or below the original Goldreich–Julian value $n_\mathrm{GJ}$.

The pulsar wind zone refers to the region that lies beyond the light cylinder, a critical boundary in pulsar magnetospheres defined by the radius
$R_{LC} = c / \Omega$, where $c$ is the speed of light and $\Omega$ denotes the angular rotation frequency of the neutron star. At this radius, the linear velocity required for magnetic field lines to co-rotate with the star reaches the speed of light, rendering rigid co-rotation physically untenable. Consequently, magnetic field lines open up, allowing a relativistic outflow of electron-positron pairs (with trace ions) that carries the pulsar's rotational energy as Poynting flux \citep{Michel91}. 
In general, the pulsar wind zone exhibit very low plasma beta $\beta$ (where $\beta$ is the ratio of thermal pressure to the magnetic pressure) due to magnetic pressure overwhelmingly dominating thermal pressure.  
This results in a plasma dominated by electromagnetic forces, making the plasma ideal for wave phenomena governed by inertia and kinetic effects rather than thermal processes.

Alfv\'{e}n waves in pulsar magnetospheres have been extensively studied \citep{Arons1986, Urpin2011} and are crucial for pulsar radio emission processes \citep{Urpin2012, Lyutikov2000}. Multiple mechanisms have been proposed for their excitation, including beam-driven instabilities at the anomalous cyclotron resonance, where streaming particle beams amplify Alfv\'{e}n waves in the outer magnetosphere \citep{Lyutikov2000, Lyutikov1999}, and nonlinear coupling, where Langmuir waves decay into Alfv\'{e}n waves in the pair plasma \citep{Gogoberidze2008}. Global plasma motions, such as differential rotation, can also drive Alfv\'{e}n perturbations through magnetorotational instability in the force-free magnetosphere \citep{Urpin2012}, providing additional launch channels. Once launched, Alfv\'{e}n waves may evolve into dispersive regimes at small scales; when wavelengths approach the ion gyro-radius or inertial length, they acquire a parallel electric field and become kinetic (or inertial) Alfv\'{e}n waves carrying field-aligned currents \citep{VEGARelativistic2024a}, which can partially convert into other modes during outward propagation. \citet{Yuan2021}, motivated by the 2016 Vela pulsar glitch with its associated radio anomaly, showed that an Alfv\'{e}n pulse launched from the neutron star’s surface (starquake-like) can traverse the magnetosphere and partly convert into fast magnetosonic waves, supporting the idea that transient internal events such as starquakes or glitches inject Alfv\'{e}n disturbances that modulate pulsar emission.

Kinetic Alfvén waves (KAWs) are one of the dispersive Alfvén modes in magnetized plasmas that may play a significant role in particle acceleration \citep{LYSAKKinetic2023} as well as plasma and energy transport \citep{BELCHERLargeamplitude1971}. KAWs emerge in the kinetic regime where $m_e/m_i \ll \beta \ll 1$ (with $m_e$ and $m_i$ the electron and ion masses respectively). In this range, thermal effects become significant at kinetic scales, with dispersion governed by the ion gyroradius when the perpendicular wavenumber ($k_\perp$) satisfies $k_\perp \rho_i \gtrsim 1$ (where $\rho_i$ is the ion gyroradius). 
KAWs propagate obliquely, with $k_\perp \gg k_\parallel$ (where $k_\parallel$ is the parallel wavenumber), and exhibit a key feature: finite parallel electric fields ($E_\parallel$) arising from charge separation between electrons and ions.
In the pulsar wind zone, the magnetic field strength decreases significantly with radial distance ($r$), typically scaling as $B_0\propto r^{-1}$ in the equatorial plane or following a more complex toroidal–poloidal structure in the relativistic outflow compared to the intense fields near the pulsar surface \citep{Michel91}. This reduction in $B$, coupled with relatively stable plasma densities (enhanced by pair cascades) and temperatures, allows the plasma beta to increase into the kinetic regime, $m_{e}/m_{i}\ll\beta\ll1$. 
Under these conditions, KAW dispersion is dominated by finite gyroradius effects, rather than ideal MHD (magnetohydrodynamic). 
Consequently, the dispersion relation is modified  
with ion gyroradius $\rho_{i}$, leading to oblique propagation, finite parallel electric fields, and enhanced energy dissipation for particle acceleration. 

KAWs can evolve into nonlinear structures such as solitons \citep{singhKineticAlfvenSolitary2019a}, and quasi-periodic wave packets \citep{singhKineticAlfvenicCnoidal2024}. Theoretical analyses often employ methods like reductive perturbation theory, the Sagdeev pseudo-potential approach, and derivations of the nonlinear Schrödinger or Hirota equations to model these structures. 
KAWs have been extensively studied in e–p–i plasmas across space and astrophysical settings \citep{Kakati2000,Shukla2004,Sah2010,AhmedSah2017,ADNANSmall2016a,Khalid2020,YuLiu2021,Singh2022,Singla2024}. These studies show that even a small ion component enables the propagation of KAWs, otherwise forbidden in pure pair plasmas. 
These findings are vital for understanding energy transport and particle acceleration in environments such as pulsar magnetospheres and the interstellar medium. 
Non-Maxwellian particle distributions (e.g., $\kappa$-distribution), common in pulsar plasmas, can significantly influence soliton amplitudes and widths by altering dispersion and nonlinearity coefficients \citep{MOUSAVIDispersion2025}.

The plasma dynamics of the pulsar wind zone beyond the light cylinder are important for understanding energy transport, current-sheet evolution, and the development of localized nonlinear structures \citep{PHILIPPOVPulsar2022}. In this region, the relativistic outflow is highly structured rather than laminar, making it a natural environment for nonlinear kinetic modes. The present work focuses on this wind-zone plasma regime and demonstrates that KA solitons can arise self-consistently in localized mass-loaded filamentary sub-structures of the outflow. Soliton-based emission scenarios have been widely discussed for inner-magnetospheric pulsar plasma conditions \citep{Weatherall1998,Melikidze2000,Gil2004,Asseo2006}, whereas the present study addresses the distinct problem of KA soliton formation and propagation in the pulsar wind zone.

While KA solitons in e-p-i plasmas have been investigated in various astrophysical and space contexts, previous studies have largely focused on their theoretical properties without establishing a direct connection to observable pulsar characteristics. The present work bridges this gap by explicitly linking the nonlinear parameters of KA solitons, such as amplitude and width, to fundamental measurable pulsar quantities including spin period, spin down rate, and magnetospheric plasma composition. 
This connection enables us to evaluate how KA soliton amplitude and width depend on measurable pulsar quantities, and how the resulting nonlinear structures can modulate the local plasma density and electrostatic environment in the wind zone. In this sense, the present model provides a self-consistent link between plasma microphysics and large-scale wind variability, while a complete radiation mechanism is left for future work.

\section{Theoretical Framework}
\label{Theoretical_Framework}

\begin{figure}
\centering
\includegraphics[width=8.5cm]{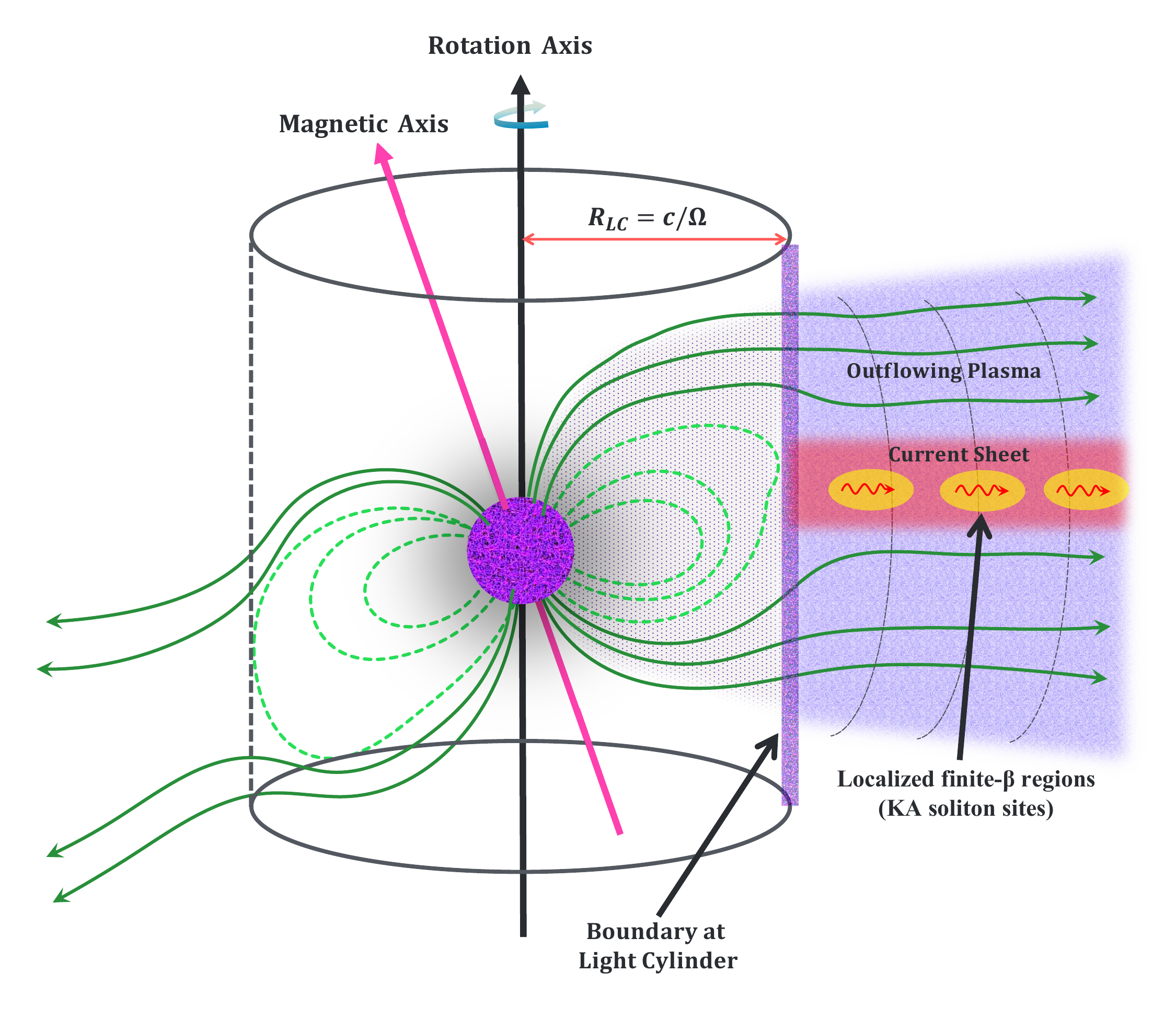}
\caption{Schematic representation of the macroscopic pulsar wind geometry beyond the light cylinder ($R_{LC} = c/\Omega$). The green field lines denote the dominant toroidal magnetic field $B_0$. Pink region represents the equatorial current sheet. Crucially, while the global relativistic wind is highly magnetized and strictly force-free ($\beta \to 0$), KA solitons (red wavy arrows) are expected to form strictly within localized, mass-loaded finite-$\beta$ regions (yellow ovals) embedded within the current sheet. These denser, lower-magnetization sub-structures provide the necessary finite inertia and weak dispersion required to support nonlinear KdV soliton dynamics.
}
\label{Final_Image}
\end{figure}

The pulsar wind zone ($r > R_{\text{LC}}$), defined by a relativistic outflow of plasma along open magnetic field lines \citep{GAENSLEREvolution2006}, presents a dynamic environment where a dense electron-positron pair plasma and a high-energy primary beam coexist \citep{ARONSPulsar2012}. 
Fig.~\ref{Final_Image} illustrates the spatial structure of the pulsar wind zone beyond the light cylinder, highlighting the magnetic-field geometry, the relativistic plasma outflow, and a current-sheet or mass-loaded region containing localized finite-$\beta$ substructures where KA solitons can form. This schematic sets the physical context for the soliton model developed in the paper. 
The streaming of this multi-component plasma provides ample free energy for wave generation,  particularly of KAWs. 
Recent particle in cell simulations have demonstrated that a resonant streaming instability, driven by the energetic pair beam, can generate strong turbulence \citep{PLOTNIKOVKinetic2024}. This turbulence spans both fluid scales, where Alfvénic modes emerge, and kinetic scales, where KAWs are excited directly \citep{VEGARelativistic2024a}. A key finding from these simulations is the development of a broad and asymmetric wave spectrum: left-handed polarized waves are strongly damped via ion-cyclotron resonance with the background ions, while right-handed modes persist and grow.
This modern kinetic perspective is corroborated by earlier analytical and hybrid models that identified multiple complementary excitation mechanisms. One of the most efficient is the anomalous cyclotron resonance, which becomes particularly effective at radii $r \gtrsim 50 R_{\text{NS}}$ (where $R_{\text{NS}}$ is the neutron star radius), allowing particles from the primary beam or energetic plasma tail to transfer energy to Alfvén waves \citep{Lyutikov2000}. Furthermore, the non-uniformity of the pulsar wind zone marked by significant gradients in density and magnetic field strength facilitates a mode conversion process, as verified by hybrid simulations \citep[e.g.,][]{Hong2012}. In this process, initial shear Alfvén waves undergo transformation into KAWs, characterized by a transition to $k_{\perp} \gg k_{\parallel}$ and the emergence of a parallel electric field $E_{\parallel}$, a hallmark of kinetic-scale physics \citep{HASEGAWAKinetic1975}.
Together, these converging lines of evidence from kinetic simulations, theoretical instability analysis, and hybrid conversion modeling provide a robust physical foundation for the generation of KAWs in the pulsar wind. This motivates the development of the following mathematical model to explore the nonlinear soliton structures arising from these interactions.

\begin{figure}
\centering
\includegraphics[width=8cm]{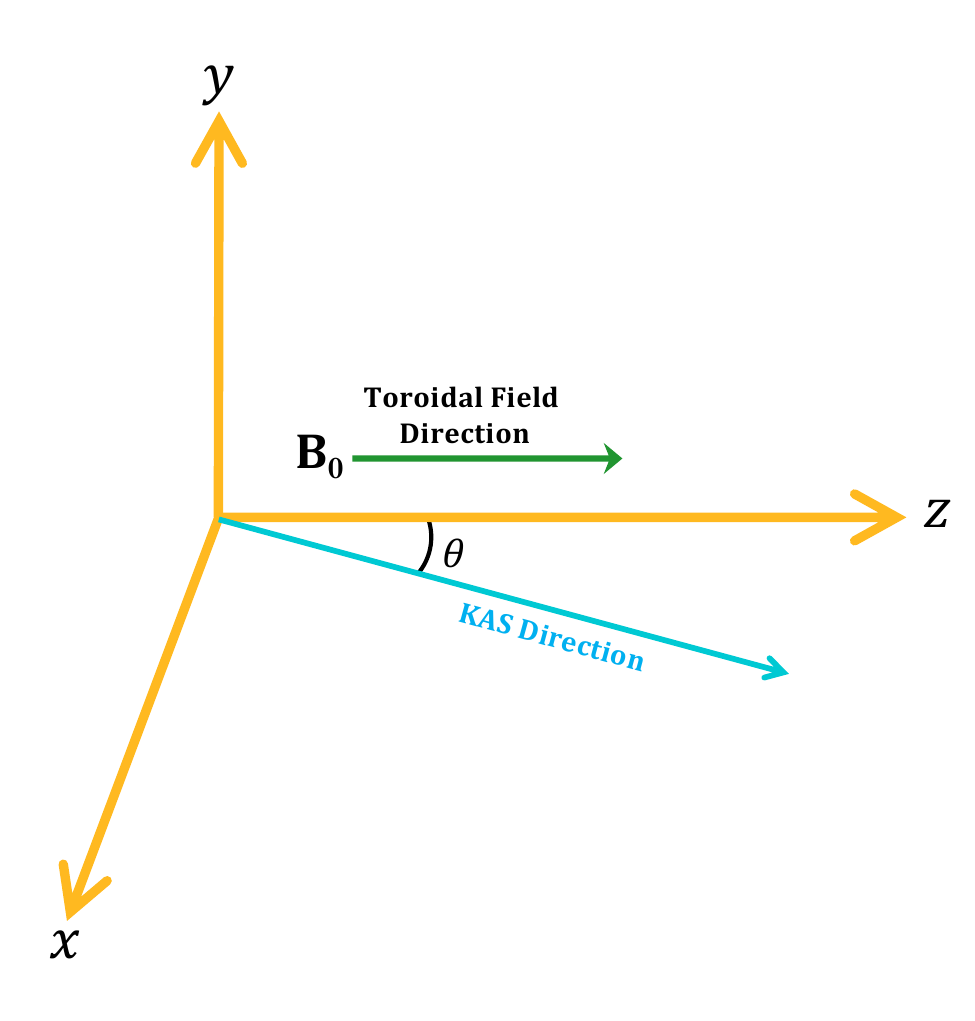}
\caption{{Local right-handed Cartesian coordinate system showing the magnetic field geometry and KA soliton (KAS) direction. The uniform background magnetic field $\bf B_0$ is aligned along the $z-$axis (toroidal direction), while the soliton propagates obliquely in the $x$-$z$ plane at angle $\theta$ to the field. This local frame is co-moving with the mass-loaded sub-structures embedded within the relativistic wind.}}
\label{geometry}
\end{figure}

Although the bulk motion of the pulsar wind is ultra-relativistic in the observer's frame of reference, the plasma can be effectively treated as non-relativistic and cold in the wind's co-moving frame. In this frame, the thermal energy of particles is negligible compared to their rest mass energy; specifically, upstream of the termination shock, the co-moving thermal energy satisfies $k_{B}T_{e,p} \ll m_e c^{2}$ for pairs and $k_{B}T_{i} \ll m_i c^{2}$ for ions, where $T_{e}$, $T_{p}$ and $T_{i}$ are the electron, positron and ion temperatures respectively. While the thermal condition allows for a non-relativistic treatment, the magnetic dynamics of the bulk wind are typically characterized by a high magnetization parameter ($\sigma\gg 1$), which would ordinarily necessitate relativistic MHD. However, our fluid description explicitly focuses on localized, high-density structures embedded within the global outflow. Modern kinetic simulations and the striped wind geometry of an oblique rotator indicate the existence of dense current sheets where the local plasma density exceeds the background Goldreich-Julian density by orders of magnitude. In these mass-loaded filaments, the local magnetization $\sigma$ is significantly suppressed, reducing the Alfvén speed sufficiently ($V_A^2 \ll c^2$) to ensure displacement currents remain negligible. Under these specific conditions, the ion component supplies the inertia that sets the KAW dispersion while the much lighter $e^{\pm}$ pairs provide the restoring force. Thus, the phase speed of KAWs remains a fraction ($\lesssim 0.3\,c$) of the speed of light, and a standard two-fluid treatment identical in form to well-tested space plasma models is adequate. Furthermore, any ultra-relativistic $e^{\pm}$ beam emerging from the polar-gap accelerator can be handled separately as a low-density driver that supplies free energy but contributes negligibly to mass loading, preserving the non-relativistic momentum balance of the background mixture.

In the equatorial pulsar wind zone beyond the light cylinder, we consider a collisionless, magnetized e--p--i plasma in which positively charged ions form a fluid medium while electrons and positrons follow the $\kappa-$distributions. In this region, we assume the dominant magnetic field component is the toroidal field $\mathbf{B}_0 =B_0\hat{z}$ (see Fig.~\ref{geometry}) produced by stellar rotation. The radial (poloidal) component \( \mathbf{B}_r \), which would act as a guide field, is significantly weaker (\( B_r / B_0 \lesssim 0.1 \)) in this region, and thus its contribution to the Lorentz force is much weaker. For analytical tractability and without loss of generality, we neglect the guide field in the present formulation. This assumption allows a closed-form derivation of the KdV equation while capturing the essential soliton physics. Wave propagation is restricted to the $x$-$z$ plane. 
To describe the KA solitons, we employ the two-potential formalism \citep{Kadomtsev1965}, where the electric field is expressed as $E = -\nabla_\perp \phi - \nabla_\parallel \psi$. Here, $\phi$ represents the potential associated with the perpendicular (inductive) electric field, while $\psi$ accounts for the parallel electrostatic potential arising from finite parallel electron thermal pressure. This approximation is valid in the low-$\beta$ kinetic regime ($m_e/m_i \ll \beta \ll 1$), where the wave phase velocity is intermediate between the electron and ion thermal velocities. Under these conditions, the nonlinear dynamics of KAWs are governed by the following normalized continuity and momentum equations respectively:

\begin{equation}
\frac{\partial n_i}{\partial t} + \nabla \cdot (n_i \mathbf{v}_i) = 0,
\end{equation}
\begin{equation}
m_i \left( \frac{\partial \mathbf{v}_i}{\partial t} + \mathbf{v}_i \cdot \nabla \mathbf{v}_i \right)
= q_i \left( \mathbf{E} + \frac{\mathbf{v}_i \times \mathbf{B}_0}{c} \right),
\end{equation}
where ${\bf v}_i=v_{ix}\hat{x}+v_{iz}\hat{z}$ is the ion fluid velocity and ${\bf E}=E_{x}\hat{x}+E_{z}\hat{z}$ is the electric field. Thus, for KA soliton propagation in $x$-$z$ plane, the ion continuity and parallel momentum equations in normalized form can thus be written as: 
\begin{equation}\label{conti}
\frac{\partial n_i}{\partial t} 
+ \frac{\partial (n_i v_{ix})}{\partial x}
+ \frac{\partial (n_i v_{iz})}{\partial z} = 0,
\end{equation}
\begin{equation}\label{moment}
\frac{\partial v_{iz}}{\partial t} 
+ v_{ix}\frac{\partial v_{iz}}{\partial x}
+ v_{iz}\frac{\partial v_{iz}}{\partial z}
= -\,\Lambda\,\frac{\partial \psi}{\partial z},
\end{equation}
where $\Lambda = Z_i \beta / 2$, and $Z_i$ denote the ion charge number.
The perpendicular ion velocity includes the \textit{polarization drift}, which arises from the time variation of the perpendicular electric field. Following the standard derivation (see Appendix in \citet{NimarKAW2016}), the polarization drift velocity in normalized form is written as:
\begin{equation}\label{drift}
v_{ix} = -\Lambda\,\frac{\partial^2 \phi}{\partial x\,\partial t}.
\end{equation}
This term couples the ion fluid motion to the perpendicular electric field perturbation and is responsible for the dispersive nature of KAWs.

Combining Ampere and Faraday's laws (see e.g., \citet{MAHMOODFully2002}), we obtain
\begin{equation}\label{maxwell}
\Lambda\frac{\partial^{4}(\phi-\psi)}{\partial x^{2}\partial z^{2}}= \frac{\partial^{2} n_{i}}{\partial t^{2}}+\frac{\partial^{2}(n_{i}v_{iz})}{\partial z \partial t}, 
\end{equation}
The Eqs.~(\ref{conti}-\ref{maxwell}) are normalized using the following characteristic scales, where primed quantities denote dimensional physical variables before normalization: the number densities are normalized by their equilibrium values, $n_j={n_j'}/{n_{j0}}$ $(j=i,e,p)$; wave potentials by the thermal potential, $(\psi,\phi)={e(\psi',\phi')}/{K_{B}T_e}$; time by the inverse ion cyclotron frequency, $t=t'\,\Omega_i$; spatial coordinates by the ion inertial length $d_i \equiv V_A/\Omega_i$, such that $(x,z)={(x',z')}/{d_i}$; and fluid velocity by the Alfvén speed, $v_{i(x,z)}={v_{i(x,z)}'}/{V_A}$. Here, $V_A=B_0/\sqrt{4\pi n_{i0}m_i}$ denotes the Alfvén velocity and $\Omega_i=eB_0/m_{i}c$ represents the ion cyclotron frequency.
The equilibrium charge neutrality condition is expressed as
\begin{equation}\label{eq5}
\mu_{ei}=1+\mu_{pi},
\end{equation}
where $\mu_{ei}={n_{e0}}/{n_{i0}},$ $\mu_{pi}={n_{p0}}/{n_{i0}}.$
The normalized $\kappa-$distributed electron ($n_e$) and positron ($n_p$) number densities in expanded form are given respectively as,
\begin{equation}\label{eq6}
n_{e}=\left[1-\frac{\psi}{(\kappa_{e}-\frac{3}{2})}\right]^{-\kappa_{e}+\frac{1}{2}}\approx 1+c_{1}\psi+c_{2}\psi^2+...
\end{equation}
where, $c_{1}=\frac{(\kappa_{e}-\frac{1}{2})}{(\kappa_{e}-\frac{3}{2})}$ and $c_{2}=\frac{(\kappa_{e}^2-\frac{1}{4})}{2(\kappa_{e}-\frac{3}{2})^2}$.
\begin{equation}\label{eq7}
n_{p}=\left[1+\frac{\alpha\psi}{(\kappa_{p}-\frac{3}{2})}\right]^{-\kappa_{p}+\frac{1}{2}}\approx 1-d_{1}\psi+d_{2}\psi^2+...
\end{equation}
where, $d_{1}=\alpha\frac{(\kappa_{p}-\frac{1}{2})}{(\kappa_{p}-\frac{3}{2})}$, $d_{2}=\alpha^2\frac{(\kappa_{p}^2 -\frac{1}{4})}{2(\kappa_{p}-\frac{3}{2})^2}$, and $\alpha={T_{e}}/{T_{p}}$ is the temperature ratio of electrons to positrons.

\section{Derivation of the KdV equation\label{derivation-kdv}}
The KdV equation is derived using the reductive perturbation method, with the independent variables $\xi$ and $\tau$ introduced through the stretched coordinates in Eqs.~(\ref{conti})--(\ref{maxwell}):
\begin{equation}\label{eq8}
\tau=\epsilon^{\frac{3}{2}}t, \quad \xi=\epsilon^\frac{1}{2}(l_{x}x+l_{z}z-\lambda t),
\end{equation}
where $\lambda$ is the phase velocity of the KAWs, $\epsilon$ is a small ($0<\epsilon<1$) parameter which denotes weak nonlinearity in the system, and $l_{x}\,(l_{z})$ are the direction cosines in the $x$($z$)-direction which can be expressed as $l_x=\rm{sin}\,(\theta)$, $l_z=\rm{cos}\,(\theta)$, with $\theta$ being the angle between the ambient magnetic field and the KAW propagation direction. 
The dependent variables can be written in the form of a power series:
\begin{equation}\label{series}
    S=\sum_{q=1}^\infty \epsilon^qS^{(q)}, \quad \rm{and} \quad \phi=\sum_{q=1}^\infty \epsilon^{q-1}\phi^{(q)},
\end{equation}
where $S=(v_{ix},v_{iz},\psi)$. The plasma approximation in the normalized form yield 
\begin{equation}\label{eq14}
n_i=\mu_{ei}n_e-\mu_{pi}n_p.
\end{equation}

\noindent
Using Eqs.~(\ref{eq6}) and (\ref{eq7}) in  Eq.~(\ref{eq14}), we find
\begin{equation}\label{eq15}
n_i=1+a_{1}\psi+a_{2}\psi^2+...
\end{equation}
where $a_1=\mu_{ei}c_1+\mu_{pi}d_{1}$ and $a_2=\mu_{ei}c_2-\mu_{pi}d_{2}$.

Using Eqs.~(\ref{eq8}), (\ref{series}), and (\ref{eq15}) in  Eqs.~(\ref{conti})--(\ref{maxwell}) and comparing the coefficients of lowest powers of $\epsilon$, we obtain the following first order equations:
\begin{equation}\label{eq16}
-a_{1} \lambda \frac{\partial \psi^{(1)}}{\partial \xi}+l_{x} \frac{\partial v_{ix}^{(1)}}{\partial \xi}+l_{z} \frac{\partial v_{iz}^{(1)}}{\partial \xi}=0,
\end{equation}
\begin{equation}\label{eq17}
v_{ix}^{(1)}= \Lambda \lambda l_{x} \frac{\partial^{2} \phi^{(1)}}{\partial \xi^{2}},
\end{equation}
\begin{equation}\label{eq18}
\lambda \frac{\partial v_{iz}^{(1)}}{\partial \xi}= \Lambda l_{z} \frac{\partial \psi^{(1)}}{\partial \xi},
\end{equation}
\begin{equation}\label{eq19}
\Lambda\, l_{x} l_{z} \frac{\partial^{2} \phi^{(1)}}{\partial \xi^{2}}= a_{1} \lambda^{2} \psi^{(1)}- l_{z} \lambda\, v_{iz}^{(1)}.
\end{equation}
Combining and simplifying Eqs.~(\ref{eq16})--(\ref{eq19}), we obtain the following expression
\begin{equation}\label{eq20}
a_{1}\lambda^{4}-(a_{1}+{\Lambda})l_{z}^{2}\lambda^{2}+{\Lambda} l_{z}^{4}=0.
\end{equation}
On solving the above biquadratic equation in $\lambda$, we obtain two different roots corresponding to KA mode and ion acoustic mode which are given respectively as:
\begin{equation}\label{eq21}
\lambda^2=l_{z}^{2}, \quad \lambda^2=\frac{\Lambda l_{z}^{2}}{a_1}.
\end{equation}

\noindent
The next order of $\epsilon$ yields the following second order evolution equations,

\begin{multline}\label{eq23}
-a_{1} \lambda \frac{\partial \psi^{(2)}}{\partial \xi}-2a_{2} \lambda \psi^{(1)} \frac{\partial \psi^{(1)}}{\partial \xi}+a_{1} \frac{\partial \psi^{(1)}}{\partial \tau}
+l_{x} \frac{\partial v_{ix}^{(2)}}{\partial \xi}\\ 
+a_{1}l_{x} \frac{\partial (\psi^{(1)}v_{ix}^{(1)})}{\partial \xi}+l_{z} \frac{\partial v_{iz}^{(2)}}{\partial \xi}
+a_{1}l_{z} \frac{\partial (\psi^{(1)}v_{iz}^{(1)})}{\partial \xi}=0,
\end{multline}
\begin{equation}\label{eq24}
v_{ix}^{(2)}= \Lambda \left(\lambda l_{x} \frac{\partial^{2} \phi^{(2)}}{\partial \xi^{2}}-l_{x}\frac{\partial^{2} \phi^{(1)}}{\partial \tau \partial \xi}\right),
\end{equation}
\begin{multline}\label{eq25}
-\lambda \frac{\partial v_{iz}^{(2)}}{\partial \xi}+\frac{\partial v_{iz}^{(1)}}{\partial \tau}+l_{x}v_{ix}^{(1)}\frac{\partial v_{iz}^{(1)}}{\partial \xi}+l_{z}v_{iz}^{(1)}\frac{\partial v_{iz}^{(1)}}{\partial \xi}
\\= -\Lambda l_{z} \frac{\partial^{2} \psi^{(2)}}{\partial \xi^{2}},
\end{multline}
\begin{multline}\label{eq26}
\Lambda l_{x}^{(2)} l_{z}^{(2)} \frac{\partial^{4} (\phi^{(2)}-\psi^{(1)})}{\partial \xi^{4}}=a_{1} \lambda^{2}\frac{\partial^{2}\psi^{(2)}}{\partial \xi^{2}}
\\+a_{2} \lambda^{2}\frac{\partial^{2}\psi^{(1)}}{\partial \xi^{2}}
-2a_{1}\lambda\frac{\partial^{2}\psi^{(1)}}{\partial \xi \partial \tau}-l_{z}\lambda \frac{\partial^{2} v_{iz}^{(2)}}{\partial \xi^{2}}\\+l_{z}\frac{\partial^{2} v_{iz}^{(1)}}{\partial \xi \partial \tau}-a_{1}l_{z}\lambda\frac{\partial^{2}(v_{iz}^{(1)}\psi^{(1)})}{\partial \xi^{2}}.
\end{multline}
\noindent
Eliminating the second order perturbed quantities from Eqs.~(\ref{eq23})--(\ref{eq26}), the following KdV equation is obtained:
\begin{equation}\label{eq27}
\frac {\partial\psi^{(1)}}{\partial \tau}+ C \psi^{(1)} \frac{\partial\psi^{(1)}}{\partial \xi}+ D \frac{\partial^{3}\psi^{(1)}}{\partial \xi^{3}}= 0,
\end{equation}
where
\begin{equation}\label{eq28}
C=-a_{1}l_{z} \quad \mathrm{and} \quad D=-\frac{l_{x}^{2}l_{z}\Lambda}{2(a_{1}-\Lambda)}
\end{equation}
are the nonlinear and the dispersion coefficients respectively. 
To derive the stationary soliton solution of the KdV equation (Eq.~\ref{eq27}), we employ a traveling wave transformation of the form $\zeta = \xi - u\tau$, where $u$ denotes the speed of the soliton. This transformation reduces the partial differential equation to an ordinary differential equation in $\zeta$, which admits an exact localized solution of the form \citep{sainiDustKineticAlfven2015}:
\begin{equation} \label{eq30}
\psi^{(1)} = \psi_0\, \mathrm{sech}^2\left(\frac{\zeta}{W}\right),
\end{equation}
where $\psi^{(1)}$ represents the leading-order perturbed parallel potential associated with the KA solitons.  Notice that $\psi_0$ and $W$ represent the amplitude and width of solitons written as
\begin{equation}
   \psi_0 = \frac{3u}{C}\quad \mathrm{and} \quad W = \sqrt{\frac{4D}{u}} \label{amplitude-width}
\end{equation}
respectively. It shows that the soliton becomes taller with increasing speed $u$ and broader with stronger dispersion $D$. This $\mathrm{sech}^2$-type solution describes a stable solitary wave structure that preserves its shape during propagation, reflecting the exact balance between nonlinear and dispersive effects in the e--p--i plasma in pulsar wind zone.

\section{Application to Pulsar Wind Zone
\label{application}}
The model developed in Sec.~\ref{Theoretical_Framework} 
applies to localized, mass-loaded filamentary 
sub-structures within the pulsar wind zone beyond 
the light cylinder, where partial magnetic-field 
dissipation produces the finite-$\beta$ conditions 
required for dispersive KAW dynamics 
(Sec.~\ref{subsec:kdv_validity}). In this 
section, the plasma parameters entering the KdV 
coefficients are expressed directly in terms of 
fundamental, observable properties of the parent 
neutron star, establishing the self-consistent 
link between soliton characteristics and pulsar 
observables.

\subsection{Plasma densities}
In the pulsar wind zone, the density of each charged species is set by the underlying Goldreich–Julian density

\begin{equation}
n_{\text{GJ}} = \frac{B_0}{e\,c\,P},
\end{equation}
where $B_0$ is the toroidal field carried by the relativistic outflow, $e$ is the elementary charge, and $P$ is the period of rotation of the neutron star. In a purely charge-separated picture, $n_{\text{GJ}}$ represents the minimum plasma density required to screen the induced electric field and enforce co-rotation; in reality, two additional processes load the wind with particles far in excess of this baseline.
First, magnetospheric cascades inject electron–positron pairs at a multiplicity of $\delta$, so that the local baseline pair densities can be written as
\begin{equation}
n_{e0} =\delta\,n_{\text{GJ}} \quad \mathrm{and} \quad n_{p0} =\delta\,n_{\text{GJ}}.
\end{equation}

\noindent
Physically, $\delta$ encapsulates the efficiency of curvature or inverse Compton induced $\gamma-$rays in the acceleration gaps above the polar caps. 
At the same time, ions stripped from the stellar surface or entrained from ambient plasma contribute an additional density 
\begin{equation}\label{ion-eta}
n_{i0} = \eta\,n_{GJ},
\end{equation}
where the ion-loading fraction $\eta$ may range from nearly unity (if the outflow is ion-dominated) down to $\ll 1$ in pair-rich winds.
As the pulsar wind expands into the nebula, both ion and pair densities decrease approximately as $1/r^2$ due to spherical divergence. Accordingly, in the wind zone, the equilibrium number densities of pairs ($n_{e0,p0}$) and ions ($n_{i0}$) can be expressed as:

\begin{eqnarray}
    n_{e0,p0}(r) &=& \delta\, n_{\rm{GJ}}\big|_{r=R_{\rm{LC}}} \left( \frac{R_{\rm{LC}}}{r} \right)^2 \nonumber \\
    &=& \delta\, \frac{B_s}{e c P} \left( \frac{R_{\rm{NS}}}{R_{\rm{LC}}} \right)^3 \left( \frac{R_{\rm{LC}}}{r} \right)^2,
\end{eqnarray}

\begin{equation}
    n_{i0}(r) = \eta\, \frac{B_s}{e c P} \left( \frac{R_{\rm{NS}}}{R_{\rm{LC}}} \right)^3 \left( \frac{R_{\rm{LC}}}{r} \right)^2,\label{ni0-wind}
\end{equation}
\noindent
where 
\begin{equation}
{n_{\rm{GJ}}}\bigr|_{r=R_{LC}}=\frac{B_{\rm LC}}{ecP}=\frac{B_s}{ecP}\left(\frac{R_{\rm{NS}}}{R_{\rm LC}}\right)^3 
\end{equation}  
is the Goldreich–Julian density evaluated at the light cylinder, with $B_{\rm{LC}} = B_s \left( R_{\rm{NS}} / R_{\rm{LC}} \right)^3$ denoting the magnetic field strength at $R_{\rm{LC}}$, and $R_{\rm{NS}}$ the neutron star radius.
Here, $B_{\rm{s}}$ is the magnetic field strength at the stellar surface. The surface field is not a free parameter but is estimated directly from the pulsar's period ($P$) and its spin-down rate ($\dot{P}$) \citep{Lyne2012}:
\begin{equation}
B_{\rm{s}} \approx 3.2 \times 10^{19} \sqrt{P \dot{P}} \,\,\rm{[Gauss]}.
\end{equation}

Since ions and pairs share the same $1/r^2-$scaled base density, their ratio
\begin{equation}
\mu_p = \frac{n_{p0}}{n_{i0}} = \frac{\delta}{\eta}
\end{equation}
remains constant throughout the wind zone. 
This single parameter governs not only the relative inertia of the plasma but also the degree of charge separation carried into the nonlinear dynamics of KA solitons. Charge neutrality then demands that the electron density exceed the ion density by one positron per ion, so that $n_{e0} = (1+\mu_p)\,n_{i0}$. 
In practice, a Crab-like pulsar with $\delta \sim 10^5$ and $\eta \sim 10^{-2}$ yields $\mu_p \sim 10^7$, indicating an overwhelmingly pair-dominated flow. Conversely, a more quiescent object with modest cascade activity might have $\delta \sim 10^2$ and $\eta \sim 10^{-1}$, giving $\mu_p \sim 10^3$.

Within our KdV framework this positron-to-ion ratio enters directly into the nonlinear coefficient $C$ and dispersive coefficient $D$ via $a_1$
\begin{equation}\label{a1}
a_1 = \bigg(1+\frac{\delta}{\eta}\bigg)\,c_1 + \frac{\delta}{\eta}\,d_1.
\end{equation}
Because $a_1$ scales linearly with $\mu_p$, when $\mu_p \gg 1$, the amplitude and width of KA solitons are extremely sensitive to the underlying pair-loading of the wind.

\subsection{Magnetic Field Model}
The structure of a pulsar's magnetic field can be understood by dividing it into two distinct regions separated by the light cylinder.
Inside the light cylinder, the magnetic field is dominated by the neutron star's intrinsic dipole. This region is characterized by closed field lines that co-rotate with the star. The magnetic field strength decreases rapidly with distance ($r$) from the star's center roughly according to the dipole formula \citep{GOLDREICHPulsar1969}:
\begin{equation}
B_0 = B_{\rm{s}} \left( \frac{R_{\rm{NS}}}{r} \right)^3
\end{equation}

\noindent
Beyond the light cylinder, the field lines can no longer co-rotate and are forced open, transitioning from a dipolar to a predominantly toroidal \emph{wind} geometry. The magnetic field strength in this region weakens much more slowly. To ensure a smooth transition at the boundary $r = R_{\text{LC}}$, the wind-zone field is modeled as decaying according to $1/r$:
\begin{equation}
B_0 = B_{\text{LC}} \frac{R_{\text{LC}}}{r}.
\end{equation}
This provides a complete and continuous model for the magnetic field across both regions:

\begin{equation}\label{final-magnetic}
B_0 = 3.2\times 10^{19}\sqrt{P\dot{P}} \left( \frac{R_{\text{NS}}}{R_{\text{LC}}} \right)^3 \frac{R_{\text{LC}}}{r}.
\end{equation}
\noindent
Thus, using the Eqs.~(\ref{a1}) and (\ref{final-magnetic}), the nonlinear and dispersive coefficients becomes, respectively

\begin{equation}\label{A}
C=-l_z \bigg[ \bigg(1+\frac{\delta}{\eta}\bigg)\,c_1 + \frac{\delta}{\eta}\,d_1 \bigg],
\end{equation}

\begin{equation}\label{finalB}
D=-\frac{l_{x}^{2}l_{z}\Lambda}{2(a_{1}-\Lambda)},
\end{equation}

\noindent
where the dimensionless parameter $\Lambda$ can be expressed in expanded form as::

\begin{equation}
    \Lambda=\frac{4 \pi Z_i\eta k_B T_e }{3.2\times 10^{19}\, e c P^{3/2} \dot{P}^{1/2}}  \left(\frac{R_{\rm{LC}}}{R_{\rm{NS}}}\right)^3.
\end{equation}

\noindent
Thus, the properties of KA solitons, such as their amplitude and width, are directly dependent on pulsar observable properties, as the nonlinear coefficient $C$ and dispersive coefficient $D$ are governed by pulsar parameters including the period ($P$), spin-down rate ($\dot{P}$), neutron star radius ($R_{\text{NS}}$) etc., which influence the magnetic field strength, plasma beta, and wind composition. Therefore, we achieve a fully self-consistent model in which every soliton property ultimately traces back to the neutron star’s observable parameters.

\subsection{Force–free limit}
\label{subsec:FF}

In relativistic pulsar winds the magnetization parameter $\sigma$ and Alfv\'{e}n velocity can be written as:
\begin{equation}
\sigma \equiv \frac{B_0^2}{4\pi\, w\, \gamma^2 c^2}, 
\qquad 
V_{A,rel} = c\,\sqrt{\frac{\sigma}{1+\sigma}}, 
\end{equation}
where $w$ is the relativistic enthalpy density defined by
\begin{equation}
w \equiv \rho c^2 + \Gamma p,
\qquad 
\rho \equiv \sum_{j} n_j m_j ,
\end{equation}
with $n_j$ and $m_j$ the proper number density and mass of species $j\in\{e,p,i\}$, $p$ the total proper pressure, and $\Gamma$ the adiabatic index. The force–free (FF) limit corresponds to $\sigma\to\infty$, where plasma inertia is negligible and $\mathbf{E}\!\cdot\!\mathbf{B}=0$.
Our KA soliton stems from the KdV equation with coefficients already given in Eqs.~(\ref{A}) and (\ref{finalB}), with \emph{$a_1$ independent of $\beta$}. Hence $C$ is $\mathcal{O}(1)$ as $\beta\to 0$, while $D$ scales with $\beta$:
\begin{equation}
D = -\,\frac{l_x^2 l_z}{4}\,\frac{\beta}{a_1-\beta/2}\nonumber
\end{equation}
\begin{equation}
= -\,\frac{l_x^2 l_z}{4 a_1}\,\beta\!\left[1+\frac{\beta}{2a_1}+\mathcal{O}(\beta^2)\right] \xrightarrow[\beta\to 0]{} 0 .
\end{equation}
Since the solitary wave solution (Eq. \ref{eq30}) has amplitude and width
\begin{equation}
\psi_0=\frac{3u}{C}, \qquad W=\sqrt{\frac{4D}{u}},
\end{equation}
respectively, therefore, for fixed $u$ and geometry/composition ($l_x,l_z,\delta,\eta$), $\psi_0$ is independent of $\beta$, and $W \propto \beta^{1/2}$.
As $\sigma\to\infty$ ($\beta\to 0$), dispersion vanishes ($D\to 0$), $W\to 0$, and the KdV balance cannot be maintained; localized KA solitons smoothly reduce to FF Alfvénic disturbances with $E_\parallel\to 0$. Thus, in practice, KA solitons are not expected in strictly FF regions but in finite–$\sigma$ pockets of the wind where mass/ion loading, current–sheet dissipation, or shear/turbulence reduce $\sigma$ and allow small yet nonzero $\beta$ and parallel electric fields.

\subsection{Self-Consistency of the KdV Ordering}
\label{subsec:kdv_validity}

The discussion above implies that KA solitons, if present in the pulsar wind, must be confined to localized finite-\(\sigma\) sub-structures rather than the force--free background flow. This naturally raises the question of whether the reductive perturbation scheme used to derive the KdV equation remains valid under such conditions. To address this, we now examine the self-consistency of the ordering assumptions in the relevant wind sub-structures. The key requirements are that the local plasma \(\beta\) remains small but finite, the perturbation amplitude satisfies \(\epsilon \ll 1\), damping does not suppress the nonlinear structure, and the resulting wave packet remains coherent over scales larger than the microscopic plasma length. 
In the following subsections, we evaluate these requirements within the model and identify the parameter regime in which they can be satisfied self-consistently.

\subsubsection{Region of applicability}

The force--free analysis of Sec.~\ref{subsec:FF} provides the basic physical boundary of applicability for the present model. In the cold, magnetically dominated inter-stripe wind, \(\beta \to 0\), so the dispersive coefficient \(D\to 0\) and the soliton width \(W\to 0\); under such conditions, the KdV balance cannot be maintained and KA solitons are not expected. At the opposite extreme, the reconnection core and X-points are strongly nonlinear and rapidly varying, so the weak-perturbation expansion underlying the reductive perturbation method is no longer valid. The relevant regime therefore lies between these two limits: localized, mass-loaded filamentary sub-structures embedded within the wind zone, where plasma compression and partial magnetic-field relaxation can produce a small but finite local \(\beta\) while the perturbation amplitude remains weak. These are the finite-\(\sigma\), weakly perturbed pockets to which the present KdV model is restricted. In what follows, we quantify this admissible regime and show that the required ordering conditions can be satisfied self-consistently within the model.

\subsubsection{Plasma beta:}
\label{subsub:beta}

The local plasma $\beta$ in the target sub-structures is estimated from the model's density (Eq.~\ref{ni0-wind}), magnetic field (Eq.~\ref{final-magnetic}), and the characteristic electron temperature $T_e = 10^8$~K adopted throughout this paper. For the representative pulsar J1220-6318 ($P = 0.7892$~s, $\dot{P} = 7.76\times10^{-17}$) at representative location $r = 10\,R_{\rm LC}$, the background parameters give $B_0 \approx 4.8\times10^{-3}$~G and $n_{i0} \approx 4.1\times10^{-4}$~cm$^{-3}$ (with $\eta = 10^{-1}$), yielding a background plasma beta
\begin{equation}
\beta =
\frac{8\pi n_{i0}k_BT_e}{B_0^2}
\approx 6.25\times10^{-7},
\label{eq:beta_bg}
\end{equation}
which confirms that the smooth background wind is effectively force-free, with $\beta \ll m_e/m_i$.

In the mass-loaded filamentary sub-structures, local plasma compression together with partial magnetic-field relaxation can increase
the effective beta. We therefore parameterize the local value as
\begin{equation}
\beta_{\rm local} = {\cal K}\,\beta,
\label{eq:beta_local}
\end{equation}
where
\begin{equation}
{\cal K} =
\frac{n_{\rm local}}{n_{i0}}
\frac{B_0^2}{B_{\rm local}^2}
\label{eq:beta-enhance}
\end{equation}
is the combined beta-enhancement factor. Here $n_{\rm local} > n_{i0}$ and $B_{\rm local} < B_0$ represent, respectively, plasma compression and partial magnetic-field relaxation inside
the dense sub-structures. Analytical and numerical studies of relativistic current sheets in pulsar winds show that reconnection proceeds in a plasmoid-dominated manner with significant plasma compression in the reconnecting sub-layers 
\citep{UZDENSKYPHYSICAL2013,CERUTTIDissipation2017}.
In addition, relativistic reconnection simulations show that plasmoids are moderately denser than the inflowing plasma by a factor of a few \citep{SIRONIPlasmoids2016}. 
These results motivate the use of localized compressed and partially demagnetized filamentary regions superposed on the smooth wind background. However, present simulations do not uniquely determine the precise values of ${\cal K}$, the filling factor of such regions, or the weakly nonlinear fluctuation amplitude relevant for the KdV ordering. For the localized mass-loaded conditions considered here, we therefore treat ${\cal K}$ as a model-consistent enhancement factor rather than as an independently predicted simulation quantity. Values in the range ${\cal K}\sim10^3$--$10^4$ raise the representative background value in Eq.~(\ref{eq:beta_bg}) to $\beta_{\rm local}\sim6\times10^{-4}$--$6\times10^{-3}$, placing the local plasma in the lower part of the finite-beta KAW regime satisfying
\begin{equation}
\frac{m_e}{m_i} < \beta_{\rm local} < 1
\label{eq:beta_ineq}
\end{equation}
for KAW dispersion. Thus, \(\beta_{\rm local}\) refers specifically to the filament interior and not to the global wind average, which remains $\beta \ll m_e/m_i$.
The adopted local beta range should therefore be interpreted as an admissible local regime for KAW dispersion, not as a uniquely predicted property of the bulk pulsar wind.

This distinction is important for the weakly nonlinear ordering as well. The perturbation parameter $\epsilon$ adopted in this work, $\epsilon\sim0.01$--0.04, is obtained from the requirement $\delta n/n\ll1$ and therefore represents an admissible range for the reductive perturbation expansion rather than an independent physical estimate of fluctuation amplitudes in pulsar-wind substructures. In this sense, the present model identifies the restricted parameter regime in which localized reconnection-modified filaments can support weakly nonlinear KA solitons, if such regions are realized.

The soliton amplitude and width (Eq.~\ref{amplitude-width}) are governed through \(\Lambda\) (Eq.~\ref{eq28}), which in the present application is evaluated using the corresponding local filament parameters, i.e., the enhanced density and reduced magnetic field entering \(\beta_{\rm local}\). For these admissible local conditions, the dispersive coefficient \(D>0\) (Eq.~\ref{finalB}), so the KA soliton branch exists within the model framework. Space-plasma observations of KA soliton-like structures at comparable local \(\beta\) values \citep{Stasiewicz2000} show that weakly nonlinear KA soliton dynamics can occur in strongly magnetized collisionless plasmas, providing phenomenological support for the type of regime considered here.

\subsubsection{Weak nonlinearity}
\label{subsub:Weak nonlinearity}

The perturbation parameter \(\epsilon\) is the ordering parameter of the reductive perturbation method and must satisfy \(0<\epsilon\ll1\). In the present work, weak nonlinearity should therefore be understood as a \emph{condition of applicability} of the KdV model, rather than as a globally established property of the pulsar-wind current sheet. The current sheet as a whole is expected to contain strongly nonlinear regions, especially near reconnection cores and X-points, where the reductive perturbation expansion is not valid. The model is restricted to 
localized mass-loaded filamentary sub-structure within which a small-amplitude KAW perturbation can propagate on top of an already established local equilibrium.

The weak-nonlinearity admissibility constraint follows from the perturbation expansion. The ion density perturbation at leading order can be written as
\begin{equation}
\frac{\delta n_i}{n_{i0}}
\simeq a_1 \psi
\simeq \epsilon\,a_1\psi^{(1)}.
\label{eq:dn_leading}
\end{equation}
At the soliton crest, \(\psi^{(1)}=\psi_0\), and using Eq.~(\ref{amplitude-width}) together with \(C=-a_1 l_z\), one obtains
\begin{equation}
\psi_0=\frac{3u}{C}
\quad \Rightarrow \quad
a_1\psi_0=-\frac{3u}{l_z}.
\end{equation}
Therefore, the peak fractional density perturbation is
\begin{equation}
\left.\frac{\delta n_i}{n_{i0}}\right|_{\rm peak}
=
\epsilon\left|\frac{3u}{l_z}\right|.
\label{eq:dn_peak_exact}
\end{equation}
For the representative parameters used in this work, \(u\simeq -0.9\) and \(l_z=\cos\theta\) with \(\theta\sim10^\circ\)–\(55^\circ\), giving \(|3u/l_z|\sim 3\)–5. Hence
\begin{equation}
\left.\frac{\delta n_i}{n_{i0}}\right|_{\rm peak}
\sim 3\text{--}5\,\epsilon.
\label{eq:dn_peak}
\end{equation}
The weak-nonlinearity requirement \(\delta n/n\ll1\) therefore gives the order-of-magnitude admissibility condition
\begin{equation}
\epsilon \ll \frac{|l_z|}{3|u|}
\sim 0.1\text{--}0.2.
\label{eq:eps_upper}
\end{equation}

In the present work we use \(\epsilon\sim0.01\)--0.04 as representative values within this weakly nonlinear admissible regime. These values are not independently inferred fluctuation amplitudes of pulsar-wind substructures; rather, they are selected to keep the reductive perturbation expansion internally consistent. Substituting this range into Eq.~(\ref{eq:dn_peak}) gives
\begin{equation}
\left.\frac{\delta n_i}{n_{i0}}\right|_{\rm peak}
\sim 0.03\text{--}0.20,
\label{eq:dn_value}
\end{equation}
so the perturbation remains at the few-percent to \(\lesssim 20\%\) level relative to the local filament equilibrium density. Values near the upper end of this interval should be regarded as the largest admissible weakly nonlinear cases rather than as independently predicted pulsar-wind fluctuation amplitudes. Thus, the adopted \(\epsilon\) range defines a model-consistent ordering regime for the KdV calculation, not a unique physical prediction for the full current sheet or for reconnection-generated substructures.

\subsubsection{Nonlinear--dispersive balance}
The reductive perturbation scheme is formulated so that slow temporal evolution, nonlinear steepening, and dispersive corrections enter at the same perturbative order. With the stretched coordinates of Eq.~(\ref{eq8}), and the expansion $\psi=\epsilon \psi^{(1)}+\epsilon^2\psi^{(2)}+\cdots,$
the leading contributions to the evolution term, nonlinear term, and dispersive term all scale as \(\mathcal{O}(\epsilon^{5/2})\). 
Thus, at the first order of the perturbation hierarchy, nonlinear and dispersive effects enter simultaneously, and the elimination of the higher-order perturbed quantities yields the KdV equation (Eq.~\ref{eq27}). In this sense, the balance between nonlinearity and dispersion is a direct consequence of the asymptotic ordering adopted in the derivation.

For this asymptotic balance to correspond to a physically realizable plasma regime, the localized wind sub-structures must simultaneously satisfy two conditions: the perturbation amplitude must remain weak (\(\epsilon\ll1\)), so that nonlinear steepening remains perturbative, and the plasma beta must be small but finite, so that the dispersive coefficient \(D\) remains nonzero. The KdV description is therefore applicable only where these two requirements are met together. 
In the present model, these two conditions are satisfied in the localized mass loaded sub-structures as discussed in Sec.~\ref{subsub:beta} and Sec.~\ref{subsub:Weak nonlinearity}. Therefore, the nonlinear--dispersive balance required for KdV soliton formation is not assumed globally throughout the pulsar wind. Rather, it is realized specifically within these localized finite-\(\sigma\), finite-\(\beta\) sub-structures where both \(\epsilon\ll 1\) and \(m_e/m_i<\beta_{\rm local}<1\) are simultaneously satisfied. Outside this regime, for example, in the force-free background wind or in the strongly nonlinear reconnection core, the balance breaks down and the KdV description is not applicable.

\subsubsection{Damping constraints}
For the nonlinear structures to survive, damping must remain weak on the soliton-evolution timescale. Collisional damping is negligible in the collisionless pulsar-wind environment considered here. Collisionless kinetic damping of KAWs, however, depends on the parallel phase speed, propagation angle, plasma beta, temperature ratios, and the particle distribution function. Because the present model is a fluid treatment of a \(\kappa\)-distributed e--p--i plasma, these kinetic damping effects are not calculated self-consistently.

The weak-damping requirement is therefore treated as a condition of applicability of the KdV model: the present description applies only to localized filamentary sub-structures for which the nonlinear evolution time is shorter than the effective kinetic-damping time. In this sense, the derived KdV solitons represent weakly damped nonlinear structures propagating within quasi-steady finite-\(\beta\) regions of the wind.
A quantitative kinetic treatment of Landau and related collisionless damping processes in the corresponding \(\kappa\)-distributed pulsar-wind plasma is left for future work.

\section{Discussions}
\label{parametric}

In this section, we discuss the physical implications of the nonlinear KA soliton solutions derived in the previous sections. We connect the analytical framework to plasma conditions relevant to the pulsar wind zone and examine how different plasma parameters and intrinsic pulsar properties influence the soliton characteristics. We first present a systematic study of how ion composition, pair multiplicity, ion loading, propagation angle, and suprathermality affect the normalized KA soliton structure, and then relate these results to the dimensional scales expected across the pulsar population.

\subsection{Effect of different ion species}
\label{subsub:Effect of different ion species}
\begin{figure}
\centering
\includegraphics[width=8cm]{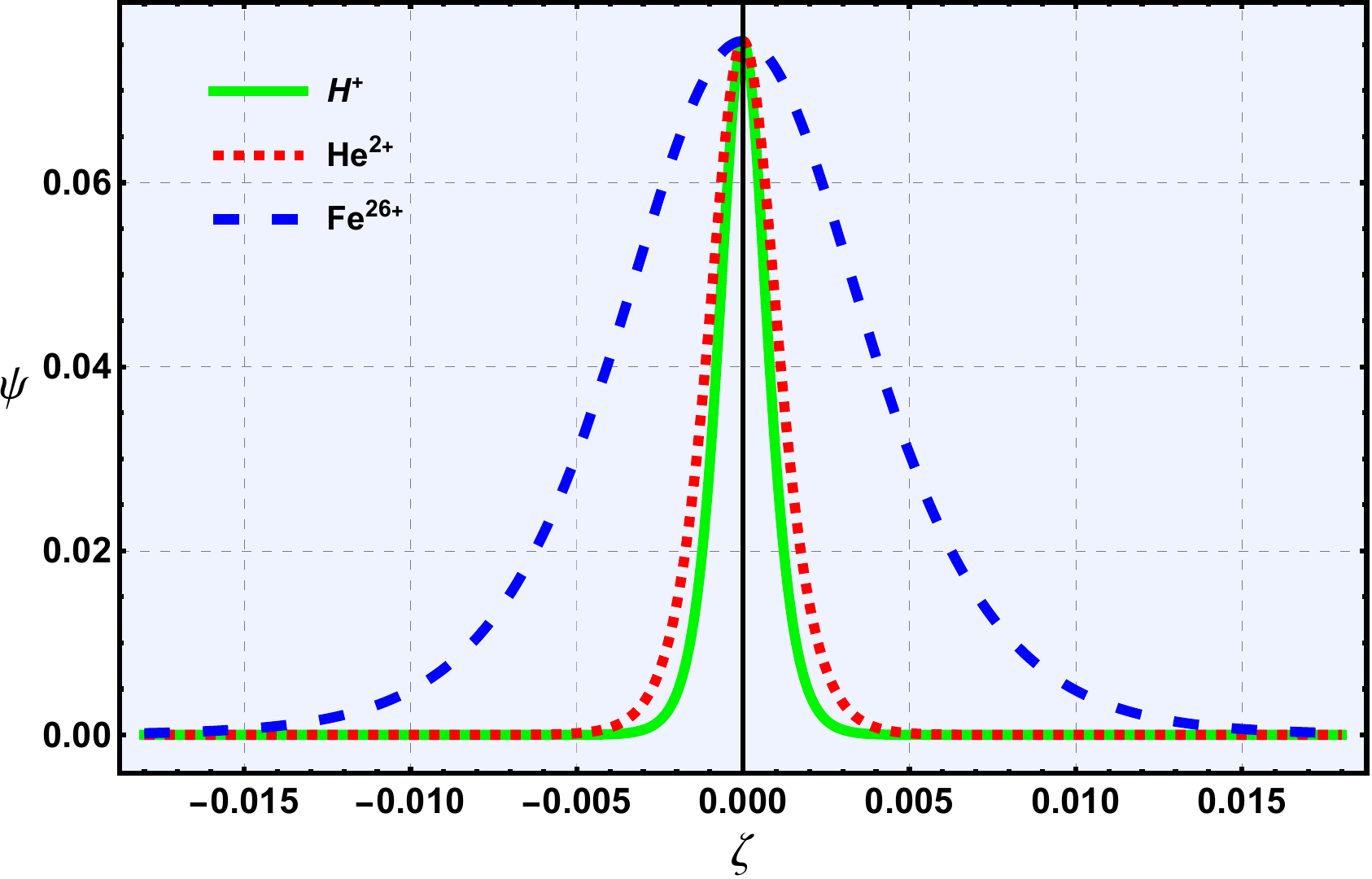}
\caption{Effect of ion species on the normalized KA soliton profile \(\psi^{(1)}(\zeta)\). The plot compares soliton structures for Hydrogen (H$^+$, green-solid), Helium (He$^{2+}$, red-dotted), and Iron (Fe$^{26+}$, blue-dashed) ions, for the pulsar \textbf{J1220-6318} with \(P=0.7892\), \(\dot P=7.760\times10^{-17}\), and keeping all other plasma parameters fixed at \(\kappa_e=\kappa_p=16\), \(\alpha=0.99\), \(\theta=55^\circ\), \(u=-0.9\), \(\delta=10^3\) and \(\eta=0.80\).}
\label{ions}
\end{figure}

The composition of ions in the pulsar wind zone can significantly impact the propagation and characteristics of KA solitons. Fig.~\ref{ions} illustrates the normalized KA soliton profile for different ion species, namely H$^{+}$, He$^{2+}$, and Fe$^{26+}$. The plot clearly demonstrates that the width of the KA solitons is highest for Fe$^{26+}$ ions (blue-dashed), followed by He$^{2+}$ (red-dotted) and then H$^{+}$ (green-solid). The presence of heavier ions, particularly highly ionized species like Fe$^{26+}$, significantly alters the inertial properties of the plasma. Heavier ions, due to their larger mass, can lead to increased effective inertia in the plasma, which in turn can influence the dispersive coefficient D of the KdV equation. A larger dispersion coefficient (D) generally results in wider solitons, as indicated by the soliton width formula in Eq. (\ref{amplitude-width}). In this comparison, the ion loading fraction is fixed at \(\eta=0.80\) for all three species, so that the differences in Fig.~\ref{ions} reflect only the effect of ion mass \(m_i\) and charge number \(Z_i\), rather than differences in ion abundance. Although the actual ion loading may in principle be species-dependent in pulsar outflows, that additional dependence is not modeled here.

In pulsar magnetospheres, particularly in the polar cap regions, the heavy ions can be extracted from the neutron star’s surface due to the extreme surface conditions, e.g., strong electric fields and high temperatures 
\citep{FAWLEYPotential1977,PROTHEROEGammaraysNeutrinos1998}. 
Theoretical models \citep{CHENGPairproduction1977,FAWLEYPotential1977} 
suggest that neutron star crusts may contain heavy elements like iron \citep{KOTERAFateUltrahigh2015}, which can be ionized and extracted by the intense electric fields induced by the pulsar’s rotation \citep{PETRIGlobalStatic2002a}.  
While protons or lighter ions (e.g., H$^{+}$, He$^{+2}$) are often assumed in simplified models, heavy ions like Fe$^{26+}$ may be more abundant in certain pulsars, especially those with strong magnetic fields
\citep{PROTHEROEGammaraysNeutrinos1998}. 
Observational evidence from X-ray spectra of some pulsars supports the presence of heavy ions, though their exact abundance remains uncertain due to complex surface chemistry and extraction processes \citep{LINXray2009}. 
The increased soliton width for Fe$^{26+}$ ions suggests that heavy ion dominated plasmas in the pulsar wind can support more stable and broader nonlinear structures than lighter ion cases.

\subsection{Pair multiplicity}
\label{subsub:Pair multiplicity}

\begin{figure}
\centering
\includegraphics[width=8cm]{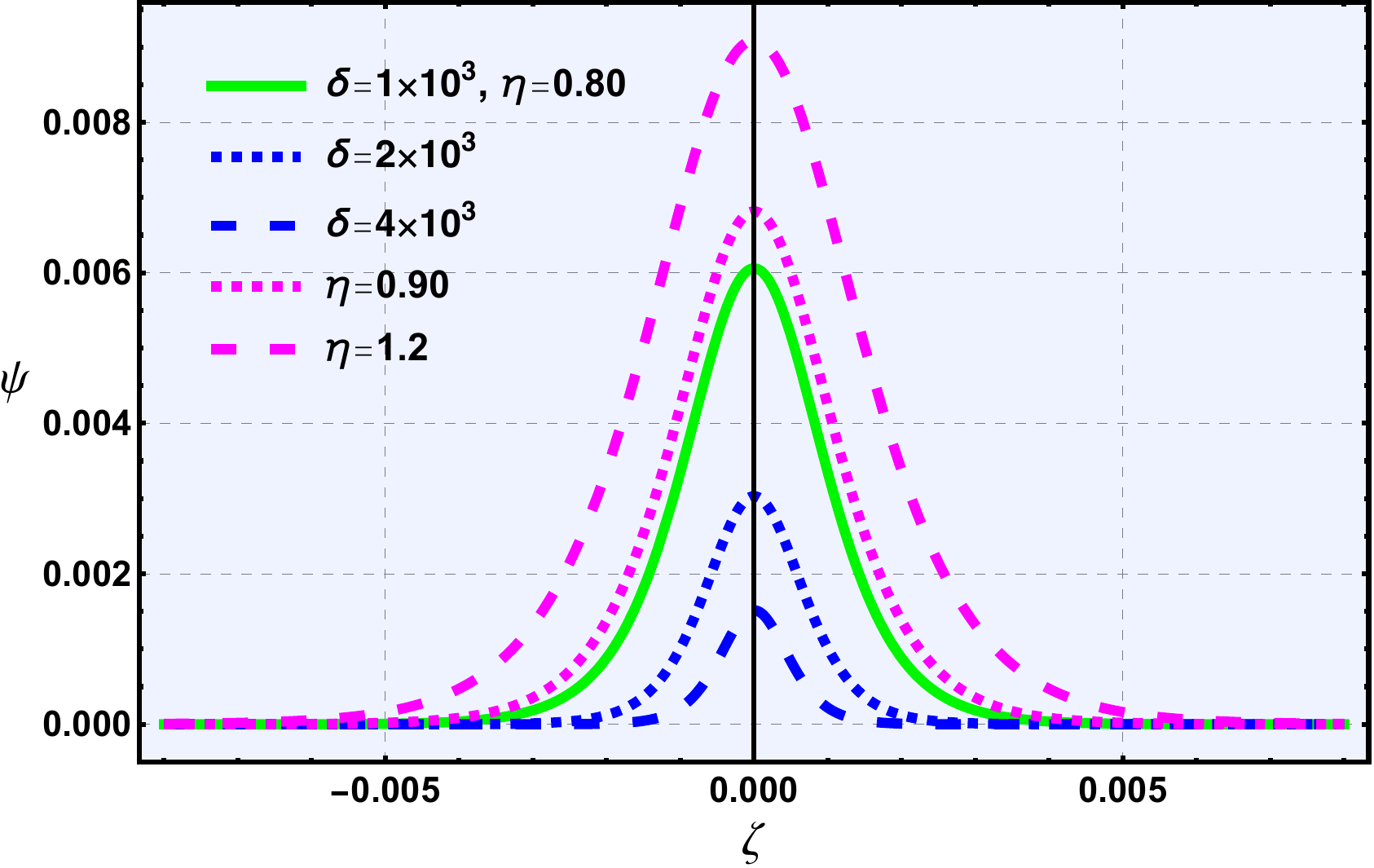}
\caption{Effect of pair multiplicity $\delta$ and ion loading factor $\eta$ on the normalized KA soliton profile $\psi^{(1)}(\zeta)$ for the same pulsar and baseline parameters as in Fig.~\ref{ions}. Only $\delta$ and $\eta$ are varied to illustrate their influence on soliton amplitude and width in an ion-rich wind ($Z_i = 26$).  Increasing $\delta$ reduces soliton amplitude and width due to enhanced pair screening, while increasing $\eta$ amplifies both via stronger inertial and dispersive effects.
}
\label{kappa-eta1}
\end{figure}

Fig.~\ref{kappa-eta1} shows that the morphology of KA solitons in pulsar winds is critically determined by the interplay between the electron-positron pair multiplicity, $\delta$, and the ion loading factor, $\eta$. An increase in pair multiplicity from $\delta=1\times10^3$ (green-solid) to $2 \times 10^3$ (blue-dotted) and $4 \times 10^3$ (blue-dashed)  leads to a reduction in both soliton amplitude and width, as a denser pair plasma dilutes the relative contribution of ions, weakens the net nonlinearity, and increases screening effects, thereby confining the wave structure to smaller spatial scales. Conversely, the soliton profile is highly sensitive to the ion loading factor, with even a modest increase in $\eta$ from $0.80$ (green-solid) to $0.90$ (magenta-dotted) and $1.2$ (magenta-dashed), resulting in a significant amplification of both soliton amplitude and width. This occurs because a higher proportion of massive ions enhances the plasma's inertial effects and magnifies the dispersive effects, which strengthens the nonlinear coupling and supports the formation of more robust, broader soliton structures.

The pair multiplicity and ion loading factor can be highly variable across different classes of pulsars \citep{SPENCERDeriving2025a}. For instance, young, energetic pulsars such as the Crab \citep{DEJAGERGammaRay1996} and Vela \citep{DEJAGERLower2007}, which have very high pair multiplicities ($\delta \sim 10^5 - 10^6$) but low ion loading ($\eta \sim 10^{-3} - 10^{-2}$), are expected to produce narrow, small amplitude KA solitons, consistent with their dynamic but weaker microstructure. However, some models \citep{TIMOKHINMaximum2019} predict the maximum achievable pair multiplicity to be few hundred thousands in pulsars with magnetic field $4\times10^{12}\, \rm{G}\lesssim B\lesssim
10^{13}\,$G and temperatures $T\gtrsim10^6$K. In the pulsar wind zone, ions represent a minority component of the plasma and can be modeled as a small fraction of the local Goldreich–Julian density, allowing their number density to be written as $n_{i0} = \eta\,n_{\text{GJ}}$, where $\eta$ is a free parameter typically much smaller than the pair multiplicity $\delta$ \citep{LYUTIKOVMassloading2003}. Despite their low number density, the large mass and charge of ions enable them to carry a dynamically significant portion of the plasma's energy and momentum, comparable to that of the relativistic pairs \citep{kirk2009pulsarwinds}.
In contrast, in some older pair-starved millisecond pulsars (MSPs), pair multiplicity can lie in the range $\delta \sim 1 - 10^3$ \citep{KISAKATeV2012a}, with a higher ion loading factor $\eta \sim 0.1 - 1$. These plasma conditions foster the generation of broader and more intense KA solitons.
These larger amplitude solitons can be more effective at accelerating particles via their parallel electric fields and creating significant density modulations.

\subsection{Angle theta}
\label{subsub:Angle theta}

\begin{figure}
\centering
\includegraphics[width=8cm]{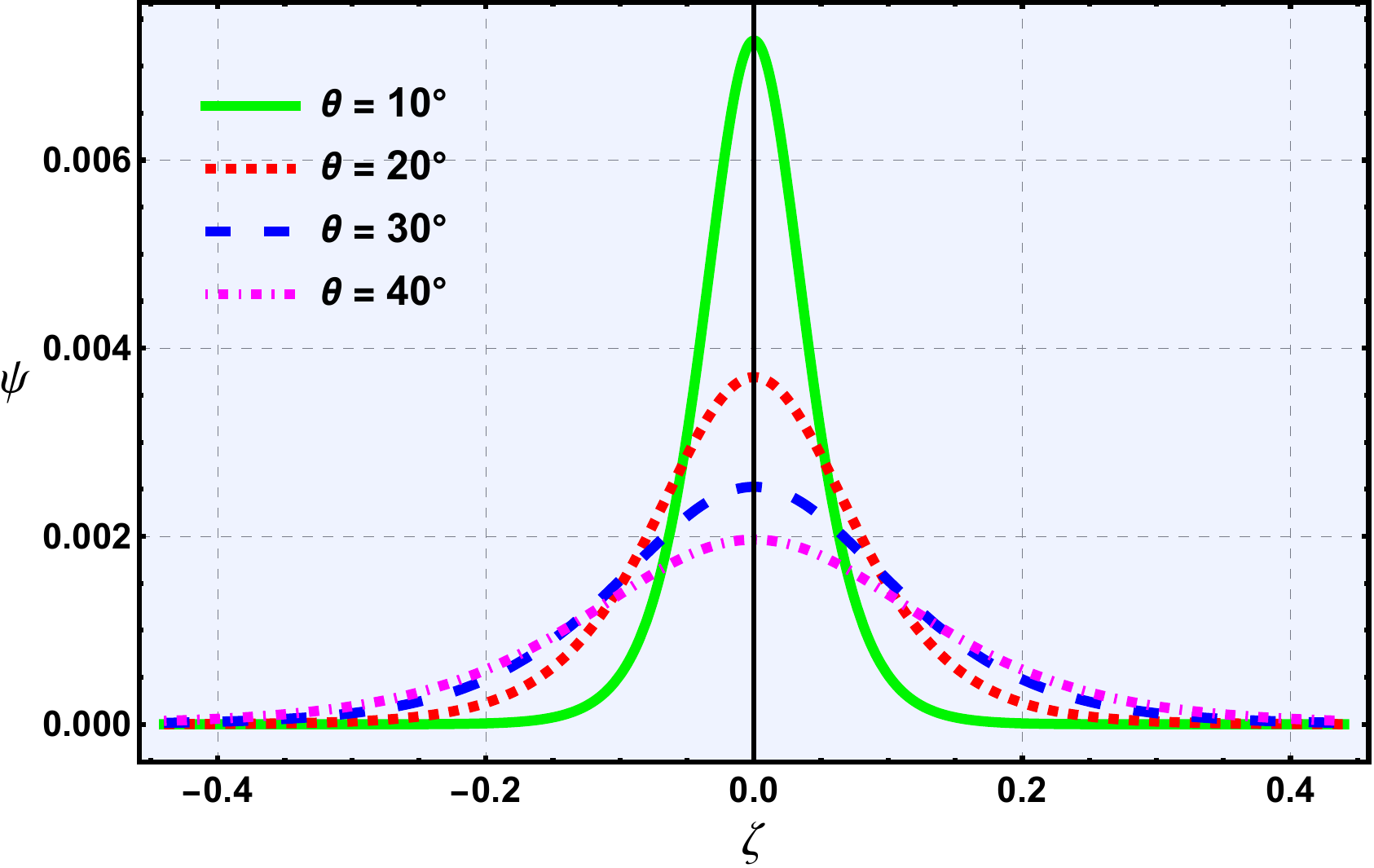}
\caption{Effect of propagation angle $\theta$ on the normalized KA soliton profile $\psi^{(1)}(\zeta)$ for the same pulsar and baseline parameters as in Fig.~\ref{ions}. Increasing $\theta$ from $10^\circ$ to $40^\circ$ leads to a reduction in soliton amplitude and an increase in width, reflecting the weakening of parallel electric fields and reduced charge bunching efficiency at larger angles.}
\label{theta-1}
\end{figure}

Fig.~\ref{theta-1} demonstrates that the angle \(\theta\) between the propagation direction of KA solitons and the mean magnetic field in the open field line region of pulsar wind significantly influence the soliton characteristics in pulsar plasmas. Specifically, an increase in \(\theta\) from $10^\circ$ (green-solid) to $20^\circ$ (red-dotted), $30^\circ$ (blue-dashed), and $40^\circ$ (magenta-dot dashed) leads to a decrease in soliton amplitude and an increase in spatial width, with the highest amplitude observed for angles close to 10$^\circ$. 
These findings can have profound implications for particle acceleration induced by KA solitons, particularly through their effects on charge separation and the motion of relativistic electron-positron pairs along magnetic field lines.

\begin{figure}
\centering
\includegraphics[width=8cm]{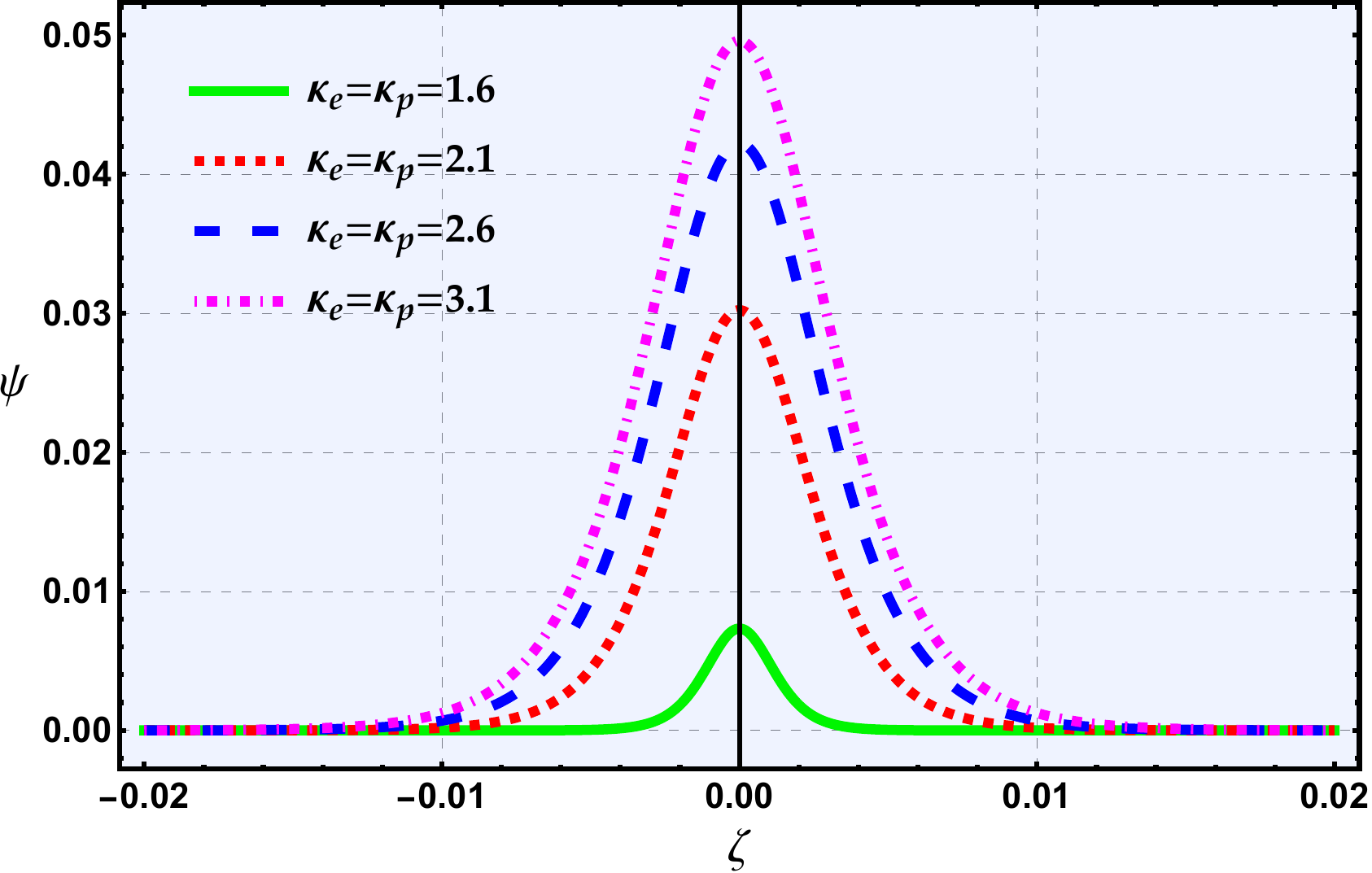}
\caption{Effect of suprathermality on the normalized KA soliton profile $\psi^{(1)}(\zeta)$ for the same pulsar and baseline parameters as in Fig.~\ref{ions}. The $\kappa$ indices for electrons and positrons ($\kappa_e = \kappa_p$) are varied to assess how nonthermal velocity distributions influence soliton structure. Higher $\kappa$ values, corresponding to more thermalized pair plasmas, lead to increased soliton amplitude and width due to reduced screening and enhanced nonlinear steepening.}
\label{kappa-fig}
\end{figure}

At small angles (\(\theta \approx 10^\circ\)), the high-amplitude solitons generate strong electric potential wells, enhancing charge separation and efficiently trapping pairs into compact bunches. These pairs, constrained by their small gyroradii to move primarily along the magnetic field, follow the field lines of the pulsar magnetosphere.  
In contrast, as \(\theta\) increases, the reduced soliton amplitude weakens the electric fields responsible for charge separation, resulting in less efficient trapping and lower pair density within the bunches. The increased soliton width further exacerbates this effect, producing larger bunches.
Additionally, broader solitons at larger angles may be more susceptible to dissipation or disruption by plasma instabilities, further limiting their ability to sustain compact bunches.

Thus, the motion of electron--positron pairs along the magnetic field, driven by the soliton's parallel electric field, implies that the nonlinear structures can create localized density and electrostatic perturbations within the wind plasma. At smaller propagation angles, the stronger and narrower solitons are expected to generate more concentrated plasma modulation, whereas at larger \(\theta\) the weaker and broader structures produce more diffuse perturbations.

\subsection{Effect of Suprathermality of pairs}
\label{subsub:Suprathermality}

The degree of suprathermality in electron and positron populations, governed by the spectral index $\kappa_{e,p}$, plays a pivotal role in shaping the properties of KA solitons within pulsar wind zone. Fig.~\ref{kappa-fig} illustrates this dependence, depicting soliton profiles for varying $\kappa_e$ and $\kappa_p$ adjusted simultaneously. A distinct trend emerges: as $\kappa$ increases from 1.6 (green-solid) to, 2.1 (red-dotted), 2.6 (blue-dashed), and 3.1 (magenta-dot dashed), both the amplitude and width of KA solitons increase significantly.
This suggests that plasmas with reduced suprathermality, approaching a Maxwellian distribution as $\kappa \to \infty$, support more taller and wider nonlinear structures.
This behavior is elucidated through the coefficients of the derived KdV equation. The soliton amplitude is inversely proportional to the nonlinear coefficient $C$ (i.e., $\psi_0 = 3u/C$), while the width scales with the square root of the dispersive coefficient $D$ (i.e., $W = \sqrt{4D/u}$). The nonlinear coefficient is directly tied to $a_1$, where $a_1 = \mu_{ei}c_{1} + \mu_{pi}d_{1}$. Parameters $c_1$ and $d_1$, functions of $\kappa_e$ and $\kappa_p$ respectively, diminish as their corresponding kappa values rise. Consequently, an increase in $\kappa_e$ and $\kappa_p$ reduces $a_1$, lowering $C$ and thus amplifying the soliton amplitude. Simultaneously, the decreased $a_1$ enhances the dispersion coefficient, resulting in a broader soliton. As suprathermality decreases and the plasma nears thermal equilibrium, the interplay of reduced nonlinearity and increased dispersion fosters taller and wider KA solitons.

Suprathermal particle distributions are a natural byproduct of intense acceleration near polar caps or magnetic reconnection sites \citep{HOSHINOSuprathermal2001}. Our findings indicate that in highly energetic regions with efficient acceleration and low-kappa (highly suprathermal) populations, KA solitons tend to be weaker and more compact. In contrast, regions with partially thermalized, higher-kappa plasmas possibly farther from acceleration sites host solitons of greater amplitude and width.

\begin{figure}
\centering
\includegraphics[width=8.25cm]{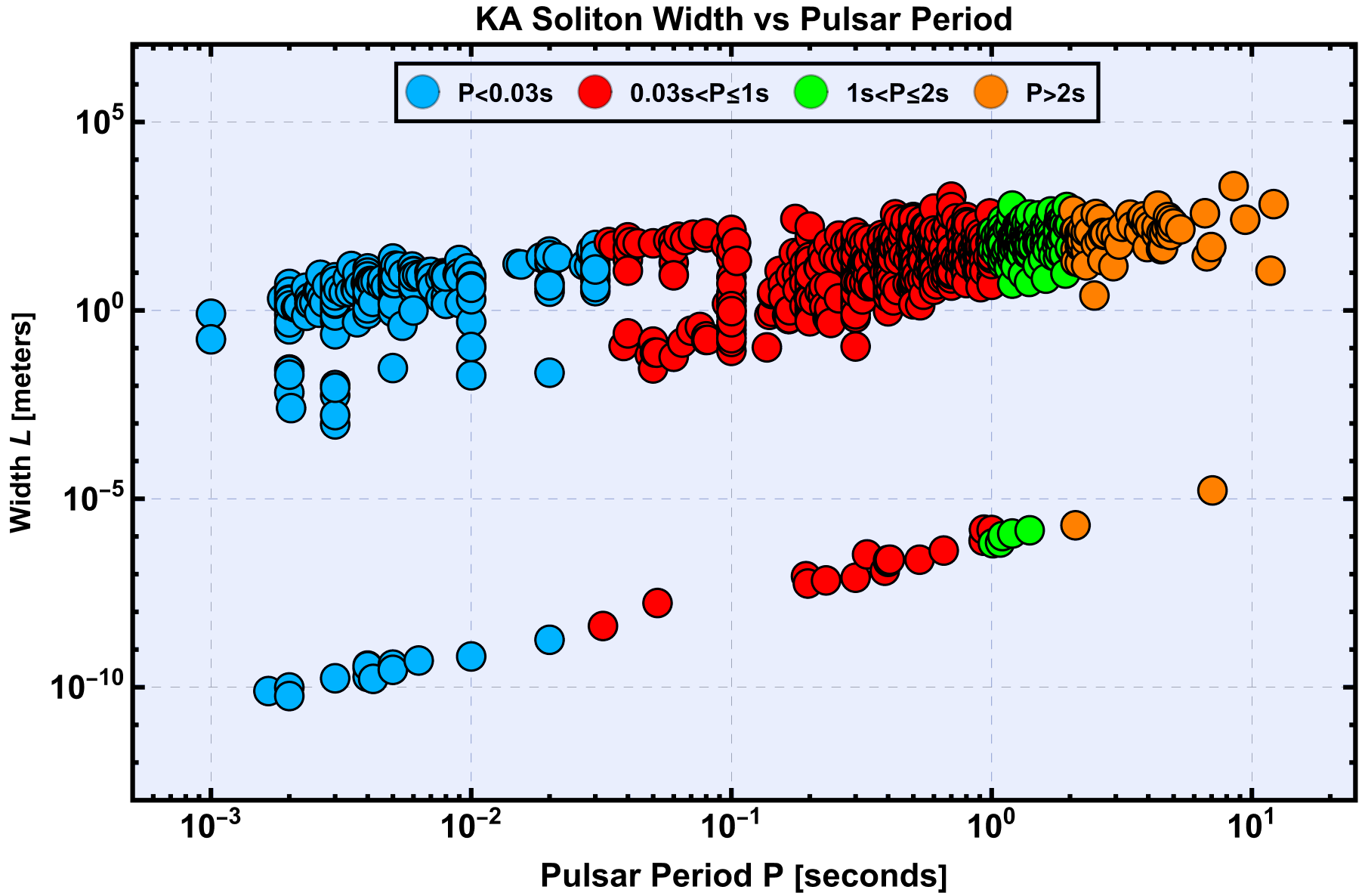}
\caption{Computed physical soliton width $L$ as a function of pulsar spin period $P$ for 1174 pulsars from the ATNF catalogue. Each point is calculated from the observed $P$ and $\dot{P}$ values using fixed plasma parameters: $\kappa_e=\kappa_p=16$, $\alpha=0.99$, $\theta=55^\circ$, $u=-0.9$, $\delta=10^3$, $\eta=0.80$, and $Z_i=26$. Millisecond pulsars (blue circles) occupy the compact end of the distribution, whereas longer-period pulsars (orange circles) support broader soliton structures. The figure illustrates how the dimensional soliton width varies systematically with the directly observed pulsar parameters $P$ and $\dot{P}$ under the adopted plasma conditions. }
\label{width-period}
\end{figure}

\subsection{Soliton Width and Pulsar Period}
\label{subsec:dimensional_width}
The KdV solution yields a normalized soliton width $W$ (Eq.~\ref{amplitude-width}). To connect this to observable pulsar parameters and produce the population-level predictions, the dimensional width must be restored.
The stretched coordinates introduced in Eq.~(\ref{eq8}) relate the physical propagation coordinates 
$s = l_x x + l_z z$ to $\xi$ by $s = \xi/\sqrt{\epsilon}$. The normalized soliton width $W$ therefore corresponds to a dimensional width
\begin{equation}
L = \frac{W\,d_i}{\sqrt{\epsilon}}.
\label{eq:Lphysical}
\end{equation}
This expression reveals that the physical  width depends not only on the plasma and  pulsar parameters encoded in $W$ and $d_i$, but also on the nonlinearity parameter $\epsilon$. 
As established in Sec.~\ref{subsub:Weak nonlinearity}, $\epsilon$ is not a free parameter: the weak-nonlinearity requirement $(\delta n/n \ll 1)$ constrains it to the range $\epsilon \sim 0.01$--$0.04$. Substituting this range into Eq.~(\ref{eq:Lphysical}) gives
\begin{equation}
L \sim (5\text{--}10)\,W\,d_i,
\end{equation}
showing that the physical soliton widths are significantly larger than the un-normalized estimate $W d_i$.

To evaluate these dimensional scales across the neutron-star population, we apply our theoretical model to 1174 pulsars from the ATNF catalogue \citep{Manchester2005}.  
Since microphysical parameters such as pair multiplicity ($\delta$) and ion loading ($\eta$) are not directly observable for individual objects, we fix these to representative theoretical values ($\delta=10^3$, $\eta=0.8$, $T_e=10^8$ K) to compute the characteristic physical scales of these structures determined by the directly measurable spin period ($P$) and spin-down rate ($\dot{P}$). Fig.~\ref{width-period} presents the calculated physical widths ($L$) for this population, assuming these structures form within the localized mass-loaded filaments described in Sec.~\ref{subsec:kdv_validity}.

The results quantify the dimensional size of these localized nonlinear structures across the pulsar population. Excluding kinematic artifacts, millisecond pulsars generally occupy the compact end of the distribution, with widths spanning approximately $10^{-3}$--$75$ m, whereas typical slow pulsars ($P>1$ s) support broader structures extending from $\sim5$ m to more than $2$\,km. These estimates provide the characteristic spatial scales over which KA solitons can modulate the local plasma density and electrostatic environment within the dense wind sub-components.

We also identify a secondary subset of data points with extremely small derived widths ($L \approx 10^{-10}$$-$$10^{-5}$\,m). These outliers correspond to pulsars with vanishingly small or kinematically contaminated measured spin-down rates, for example globular-cluster pulsars affected by the Shklovskii effect \citep{Shklovskii1970}. Since the theoretical width scales as $L \propto (P\dot{P})^{1/4}$, an anomalously small $\dot{P}$ can drive the inferred widths to extremely small values. To assess whether the resulting structures remain compatible with KAW dispersion, we compared the derived widths with the characteristic ion gyroradius $\rho_i$ obtained from the same plasma model parameters. This comparison shows that a substantial portion of the physically relevant population (in all pulsar classes) lies in the regime $L \sim \rho_i$, or more generally satisfies the finite-gyroradius condition $k_\perp \rho_i \gtrsim 1$, confirming that the dispersive requirement for KAW dynamics is met in the bulk of the population. By contrast, the extreme low-$L$ outliers fall far below the characteristic gyroradius scale and is most naturally interpreted as an artifact of the input timing parameters rather than as a population of physically meaningful wave structures. Although variations in the unobserved parameters $\delta$ and $\eta$ would introduce vertical scatter into the distribution, the overall separation of scales between MSPs and slow pulsars remains a robust consequence of the macroscopic light-cylinder geometry.

\subsection{Electrostatic Potential of KA Solitons}
We now translate the dimensionless soliton solutions into physical electrostatic scales relevant for pulsar plasma environments. The normalized amplitudes discussed in Secs.~\ref{subsub:Effect of different ion species} to \ref{subsec:dimensional_width}, can be interpreted in physical units by restoring the perturbation expansion. In the reductive perturbation framework, the normalized potential is written as $\psi=\epsilon\psi^{(1)}+\cdots$, so that the corresponding dimensional physical electrostatic potential is
\[
\psi' = \left(\frac{k_B T_e}{e}\right)\epsilon \psi^{(1)},
\]
with a peak value of
\[
\psi'_{\rm peak} = \left(\frac{k_B T_e}{e}\right)\epsilon \psi_0.
\]
For a representative electron temperature $T_e=10^8$~K, one has $k_B T_e/e \approx 8619$~V. Using the weakly nonlinear range $\epsilon\sim0.01$ to $0.04$ established in Sec.~\ref{subsub:Weak nonlinearity}, the normalized amplitudes obtained across the parameter space explored in Secs.~\ref{subsub:Effect of different ion species} to \ref{subsec:dimensional_width} correspond to sub-thermal dimensional electrostatic potentials of order
\[
\psi' \sim \mathcal{O}(0.1 \text{ to } 10^1)\ \mathrm{V},
\]
depending on plasma composition and propagation parameters.

These potentials act over the physical soliton scale $L$
derived in Eq.~(\ref{eq:Lphysical}), and therefore define a localized parallel electric field of order
$E_\parallel' \sim {\psi'}/{L}$.
Thus, the electrostatic potential should be interpreted not as an isolated voltage, but as the amplitude of a localized electrostatic structure of spatial extent $L$ embedded in the pulsar wind plasma.

The same perturbation corresponds to a fractional density variation of order
\[
\frac{\delta n}{n_0} \sim \epsilon \psi^{(1)},
\]
so the KA soliton represents a localized electrostatic density structure whose characteristic size is set by $L$. Since
\[
\frac{e\psi'_{\rm peak}}{k_B T_e} = \epsilon \psi_0 \ll 1,
\]
the perturbation remains strictly sub-thermal and is fully consistent with the weakly nonlinear ordering adopted in this work.

In a strongly magnetized pulsar wind characterized by a background magnetic field $B_0$, the ideal condition strictly enforces $E_\parallel' = 0$ along the magnetic field lines. While the energy density of the soliton's electrostatic perturbation is negligible compared to the local background magnetic energy density ($B_0^2/8\pi$), the KA soliton introduces a localized parallel electric field $E_\parallel' \sim \psi'/L$, representing a finite and critical departure from the ideal force-free constraint.

Overall, these results establish that KA solitons arising in mass-loaded filamentary sub-structures of the pulsar wind constitute self-consistent nonlinear plasma structures whose amplitude and spatial scale are determined by measurable pulsar parameters. The localized parallel electric fields and density perturbations they introduce represent a physically meaningful departure from the force-free background, with implications for plasma energy transport and the microphysics of pulsar outflows.

\section{CONCLUSION AND SUMMARY OF RESULTS}

We have developed a self-consistent theoretical framework for the formation and propagation of KA solitons in localized, mass-loaded filamentary sub-structures within the pulsar wind zone. Modeling the plasma as a collisionless, magnetized e--p--i mixture and employing reductive perturbation theory, we derived a KdV equation that governs the nonlinear evolution of these structures. Within this framework, the soliton amplitude and width are linked directly to pulsar observables through the wind density, magnetic-field strength, plasma composition, and propagation geometry.

A central result of this work is that KA solitons are not expected in the strictly force-free background wind, where \(\beta \rightarrow 0\) and the dispersive coefficient vanishes. Instead, the physically relevant regime is confined to localized finite-\(\sigma\), finite-\(\beta\) sub-structures embedded within the wind, where mass loading and partial magnetic-field relaxation permit weak but finite dispersion. We showed that, in this regime, the KdV ordering is self-consistent: the perturbation amplitude remains small, damping is negligible, the local-background approximation is well satisfied, and the nonlinear--dispersive balance required for soliton formation can be realized. The finite-\(\beta\) and weak-amplitude ranges used here should therefore be interpreted as model-consistent admissible regimes for localized wind sub-structures, rather than as uniquely predicted plasma conditions.

Our parametric analysis shows that soliton morphology depends sensitively on plasma composition and propagation geometry. Heavier ion species support broader structures through enhanced inertia and dispersion, increasing pair multiplicity suppresses the soliton scale through stronger screening, and more oblique propagation produces wider but lower-amplitude structures. Likewise, increasing the \(\kappa\) indices of the pairs, corresponding to more thermalized distributions, yields taller and broader solitons by modifying the nonlinear and dispersive coefficients. In this way, the model provides a direct connection between pulsar plasma conditions and the morphology of the resulting KA solitons.

A population-level analysis of 1174 pulsars further shows that the dimensional width of these structures varies systematically across the pulsar population. After restoring the stretched-coordinate scaling, the widths plotted in Fig.~\ref{width-period} represent physically constrained soliton scales rather than arbitrary normalization. Excluding kinematic artifacts, MSPs occupy the compact end of the distribution, with characteristic dimensional widths of $10^{-3}$--$75$ m, whereas typical slow pulsars support systematically broader structures of $5$\,m to $2$\,km.

The astrophysical significance of the present results lies primarily in the role of KA solitons as plasma-structuring agents in the pulsar wind. The localized density enhancements and electrostatic potential structures derived here can modulate the local plasma environment, modify charge separation at the fluid level, and potentially influence secondary variability and dissipation within the outflow. In particular, such nonlinear structures may affect current-sheet gradients and energy-conversion processes in the equatorial wind, thereby motivating future analytical and kinetic studies of soliton-assisted dissipation in striped-wind models.

Future work should therefore proceed in two complementary directions. First, particle-in-cell and hybrid kinetic simulations are needed to test whether the localized finite-\(\beta\) filamentary structures assumed here arise naturally and sustain KA-like nonlinear wave packets under realistic pulsar-wind conditions. Second, observational and theoretical studies should examine how such structures may contribute to wind-zone variability, dissipation, and plasma modulation in neutron-star outflows. In this sense, the present model provides a physically motivated starting point for connecting nonlinear wave dynamics in the pulsar wind to larger-scale plasma behavior beyond the light cylinder.

\section{Data Availability}
No observational or numerical data have been produced in this work. The results of all of the analytical calculations used in this work are presented in the text.

\textbf{Acknowledgement:} M.S. gratefully acknowledges financial support from the Basic Scientific Research Fund for Central Universities, China (Grant No. 2682025CX094). G.S. and N.S.S. acknowledge funding from the Department of Science and Technology, Government of India, under the DST-SERB project (Grant No. CRG/2019/003988). SL acknowledge the support by National Natural Science Foundation of China under the Grant No. 12375103.

\bibliography{References}{}

@article{AhmedSah2017,
  title = {Solitary Kinetic {{Alfv\'en}} Waves in a Dense Electron--Positron--Ion Plasma with Degenerate Electrons and Positrons},
  author = {Ahmed, Monzurul K and Sah, Om P},
  year = 2017,
  month = oct,
  journal = {Plasma Science and Technology},
  volume = {19},
  number = {12},
  pages = {125302},
  publisher = {IOP Publishing},
  issn = {1009-0630},
  doi = {10.1088/2058-6272/aa8765},
  urldate = {2026-04-06}
}

@article{ADNANSmall2016a,
  title = {Small Amplitude {{Kinetic Alfven}} Waves in a Superthermal Electron--Positron--Ion Plasma},
  author = {Adnan, Muhammad and Mahmood, Sahahzad and Qamar, Anisa and Tribeche, Mouloud},
  year = 2016,
  month = nov,
  journal = {Advances in Space Research},
  volume = {58},
  number = {9},
  pages = {1746--1754},
  issn = {0273-1177},
  doi = {10.1016/j.asr.2016.07.009},
  urldate = {2026-04-06}
}

@article{ARONSPulsar2012,
  title = {Pulsar {{Wind Nebulae}} as {{Cosmic Pevatrons}}: {{A Current Sheet}}'s {{Tale}}},
  shorttitle = {Pulsar {{Wind Nebulae}} as {{Cosmic Pevatrons}}},
  author = {Arons, Jonathan},
  year = {2012},
  month = nov,
  journal = {Space Science Reviews},
  volume = {173},
  number = {1},
  pages = {341--367},
  issn = {1572-9672},
  doi = {10.1007/s11214-012-9885-1},
  urldate = {2025-08-23},
  langid = {english}
}

@article{Arons83,
  author  = {J.~Arons},
  title   = {Pair creation above pulsar polar caps -- Geometrical structure and energetics of slot gaps},
  journal = {The Astrophysical Journal},
  volume  = {266},
  year    = {1983},
  pages   = {215},
}

@article{Arons1986,
  author = {Arons, J. and Barnard, J. J.},
  title = {Wave propagation in pulsar magnetospheres {I}: Dispersion relations and polarization},
  journal = {The Astrophysical Journal},
  volume = {302},
  pages = {120},
  year = {1986},
  doi = {10.1086/163978}
}

@article{Asseo2006,
  author = {Asseo, E. and Porzio, A.},
  title = {Strong {L}angmuir turbulence in a pulsar emission region: statistical analysis},
  journal = {Monthly Notices of the Royal Astronomical Society},
  volume = {369},
  pages = {1469--1490},
  year = {2006},
  doi = {10.1111/j.1365-2966.2006.10386.x}
}

@article{BELCHERLargeamplitude1971,
  title = {Large-Amplitude {{Alfv{\'e}n}} Waves in the Interplanetary Medium, 2},
  author = {Belcher, J. W. and Davis Jr., Leverett},
  year = {1971},
  journal = {Journal of Geophysical Research},
  volume = {76},
  number = {16},
  pages = {3534--3563},
  issn = {2156-2202},
  doi = {10.1029/JA076i016p03534},
  urldate = {2025-08-19},
  copyright = {Copyright {\copyright} 1971 by the American Geophysical Union.},
  langid = {english}
}

@article{Benacek2024,
  title = {Streaming Instability in Neutron Star Magnetospheres: {{No}} Indication of Soliton-like Waves},
  shorttitle = {Streaming Instability in Neutron Star Magnetospheres},
  author = {Ben{\'a}{\v c}ek, Jan and Mu{\~n}oz, Patricio A. and B{\"u}chner, J{\"o}rg and Jessner, Axel},
  year = 2024,
  month = mar,
  journal = {Astronomy \& Astrophysics},
  volume = {683},
  pages = {A69},
  issn = {0004-6361, 1432-0746},
  doi = {10.1051/0004-6361/202348087},
  urldate = {2026-04-06},
  copyright = {https://creativecommons.org/licenses/by/4.0}
}

@article{CERUTTIDissipation2017,
  title = {Dissipation of the Striped Pulsar Wind},
  author = {Cerutti, B. and Philippov, A. A.},
  year = 2017,
  month = nov,
  journal = {Astronomy \& Astrophysics},
  volume = {607},
  pages = {A134},
  publisher = {EDP},
  issn = {0004-6361},
  doi = {10.1051/0004-6361/201731680},
  urldate = {2025-10-29}
}

@article{CHENGPairproduction1977,
  title = {Pair-Production Discharges above Pulsar Polar Caps.},
  author = {Cheng, A. F. and Ruderman, M. A.},
  year = {1977},
  month = jun,
  journal = {The Astrophysical Journal},
  volume = {214},
  pages = {598--606},
  publisher = {IOP},
  issn = {0004-637X},
  doi = {10.1086/155285},
  urldate = {2025-07-17}
}

@article{DEJAGERLower2007,
  title = {Lower {{Limits}} on {{Pulsar Pair Production Multiplicities}} from {{H}}.{{E}}.{{S}}.{{S}}. {{Observations}} of {{Pulsar Wind Nebulae}}},
  author = {de Jager, O. C.},
  year = {2007},
  month = apr,
  journal = {The Astrophysical Journal},
  volume = {658},
  number = {2},
  pages = {1177--1182},
  publisher = {American Astronomical Society},
  issn = {0004-637X, 1538-4357},
  doi = {10.1086/511950},
  urldate = {2025-07-17}
}

@article{DEJAGERGammaRay1996,
  title = {Gamma-{{Ray Observations}} of the {{Crab Nebula}}: {{A Study}} of the {{Synchro-Compton Spectrum}}},
  shorttitle = {Gamma-{{Ray Observations}} of the {{Crab Nebula}}},
  author = {{de Jager}, O. C. and Harding, A. K. and Michelson, P. F. and Nel, H. I. and Nolan, P. L. and Sreekumar, P. and Thompson, D. J.},
  year = {1996},
  month = jan,
  journal = {The Astrophysical Journal},
  volume = {457},
  pages = {253},
  publisher = {IOP},
  issn = {0004-637X},
  doi = {10.1086/176726},
  urldate = {2025-07-17} 
}

@article{FAWLEYPotential1977,
  title = {Potential Drops above Pulsar Polar Caps: Acceleration of Nonneutral Beams from the Stellar Surface.},
  shorttitle = {Potential Drops above Pulsar Polar Caps},
  author = {Fawley, W. M. and Arons, J. and Scharlemann, E. T.},
  year = {1977},
  month = oct,
  journal = {The Astrophysical Journal},
  volume = {217},
  pages = {227--243},
  publisher = {IOP},
  issn = {0004-637X},
  doi = {10.1086/155573},
  urldate = {2025-07-16}
}

@article{sainiDustKineticAlfven2015,
  title = {Dust Kinetic {{Alfv{\'e}n}} Solitary and Rogue Waves in a Superthermal Dusty Plasma},
  author = {Saini, N. S. and Singh, Manpreet and Bains, A. S.},
  year = {2015},
  month = nov,
  journal = {Physics of Plasmas},
  volume = {22},
  number = {11},
  pages = {113702},
  issn = {1070-664X, 1089-7674},
  doi = {10.1063/1.4935165},
  urldate = {2025-06-24}
}

@article{GAENSLEREvolution2006,
  title = {The {{Evolution}} and {{Structure}} of {{Pulsar Wind Nebulae}}},
  author = {Gaensler, Bryan M. and Slane, Patrick O.},
  year = {2006},
  month = sep,
  journal = {Annual Review of Astronomy and Astrophysics},
  volume = {44},
  number = {Volume 44, 2006},
  pages = {17--47},
  publisher = {Annual Reviews},
  issn = {0066-4146, 1545-4282},
  doi = {10.1146/annurev.astro.44.051905.092528},
  urldate = {2025-08-23}
}

@article{Gil2004,
  author = {Gil, J. and Lyubarsky, Y. and Melikidze, G. I.},
  title = {Curvature radiation in pulsar magnetospheric plasma},
  journal = {The Astrophysical Journal},
  volume = {600},
  pages = {872--882},
  year = {2004},
  doi = {10.1086/379972},
  eprint = {astro-ph/0310621}
}

@article{HASEGAWAKinetic1975,
  title = {Kinetic Process of Plasma Heating Due to {{Alfv{\'e}n}} Wave Excitation},
  author = {Hasegawa, Akira and Chen, Lui},
  year = {1975},
  month = aug,
  journal = {Physical Review Letters},
  volume = {35},
  pages = {370--373},
  issn = {0031-9007},
  doi = {10.1103/PhysRevLett.35.370},
  urldate = {2025-08-23}
}

@article{Hong2012,
  author  = {Hong, M. H. and Lin, Y. and Wang, X. Y.},
  title   = {{Generation of kinetic Alfv{\'e}n waves by beam-plasma interaction in non-uniform plasma}},
  journal = {Physics of Plasmas},
  year    = {2012},
  volume  = {19},
  number  = {7},
  pages   = {072903},
  doi     = {10.1063/1.4736988}
}

@article{HOSHINOSuprathermal2001,
  title = {Suprathermal Electron Acceleration in Magnetic Reconnection},
  author = {Hoshino, M. and Mukai, T. and Terasawa, T. and Shinohara, I.},
  year = {2001},
  journal = {Journal of Geophysical Research},
  volume = {106},
  number = {A11},
  pages = {25979--25997},
  issn = {2156-2202},
  doi = {10.1029/2001JA900052},
  urldate = {2025-08-23},
}

@article{Kakati2000,
  title = {On the Existence of Small Amplitude Double Layers in an Electron--Positron--Ion Plasma},
  author = {Kakati, H. and Goswami, K. S.},
  year = 2000,
  month = mar,
  journal = {Physics of Plasmas},
  volume = {7},
  number = {3},
  pages = {808--813},
  issn = {1070-664X},
  doi = {10.1063/1.873875},
  urldate = {2026-04-06}
}

@article{NimarKAW2016,
  title = {Ion Acoustic Kinetic {{Alfv{\'e}n}} Rogue Waves in Two Temperature Electrons Superthermal Plasmas},
  author = {Kaur, Nimardeep and Saini, N. S.},
  year = 2016,
  month = oct,
  journal = {Astrophysics and Space Science},
  volume = {361},
  number = {10},
  pages = {331},
  issn = {0004-640X, 1572-946X},
  doi = {10.1007/s10509-016-2917-7},
  urldate = {2025-10-29}
}

@article{Khalid2020,
  title = {Alfvenic Perturbations with Finite {{Larmor}} Radius Effect in Non-{{Maxwellian}} Electron--Positron--Ion Plasmas},
  author = {Khalid, Saba and Qureshi, M. N. S. and Masood, W.},
  year = 2020,
  month = feb,
  journal = {AIP Advances},
  volume = {10},
  number = {2},
  publisher = {AIP Publishing},
  doi = {10.1063/1.5141891},
  urldate = {2026-04-06}
}

@article{KISAKATeV2012a,
  title = {{{TeV}} Cosmic-ray Electrons from Millisecond Pulsars},
  author = {Kisaka, Shota and Kawanaka, Norita},
  year = {2012},
  month = apr,
  journal = {Monthly Notices of the Royal Astronomical Society},
  volume = {421},
  number = {4},
  pages = {3543--3549},
  issn = {0035-8711},
  doi = {10.1111/j.1365-2966.2012.20576.x},
  urldate = {2025-07-18}
}

@article{KOTERAFateUltrahigh2015,
  title = {The Fate of Ultrahigh Energy Nuclei in the Immediate Environment of Young Fast-Rotating Pulsars},
  author = {Kotera, Kumiko and Amato, Elena and Blasi, Pasquale},
  year = {2015},
  month = aug,
  journal = {Journal of Cosmology and Astroparticle Physics},
  volume = {2015},
  number = {08},
  pages = {026--026},
  publisher = {IOP Publishing},
  issn = {1475-7516},
  doi = {10.1088/1475-7516/2015/08/026},
  urldate = {2025-07-11}
}

@article{LINXray2009,
  title = {The {{X-ray Properties}} of the {{Energetic Pulsar PSR J1838-0655}}},
  author = {Lin, Lupin Chun-Che and Takata, Jumpei and Hwang, Chorng-Yuan and Liang, Jau-Shian},
  year = {2009},
  month = nov,
  journal = {Monthly Notices of the Royal Astronomical Society},
  volume = {400},
  number = {1},
  eprint = {0907.5308},
  primaryclass = {astro-ph},
  pages = {168--175},
  issn = {0035-8711, 1365-2966},
  doi = {10.1111/j.1365-2966.2009.15468.x},
  urldate = {2025-07-17}
}

@book{Lyne2012,
  author    = {Andrew Lyne and Francis Graham-Smith},
  title     = {Pulsar Astronomy},
  edition   = {4},
  year      = {2012},
  publisher = {Cambridge University Press},
  address   = {Cambridge, UK},
  pages     = {71},
  isbn      = {978-1-107-01014-7},
}

@article{LYSAKKinetic2023,
  title = {Kinetic {{Alfv{\'e}n}} Waves and Auroral Particle Acceleration: A Review},
  shorttitle = {Kinetic {{Alfv{\'e}n}} Waves and Auroral Particle Acceleration},
  author = {Lysak, R. L.},
  year = {2023},
  month = jan,
  journal = {Reviews of Modern Plasma Physics},
  volume = {7},
  number = {1},
  pages = {6},
  issn = {2367-3192},
  doi = {10.1007/s41614-022-00111-2},
  urldate = {2025-08-19}
}

@article{LYUTIKOVMassloading2003,
  title = {Mass-Loading of Pulsar Winds},
  author = {Lyutikov, M.},
  year = {2003},
  month = mar,
  journal = {Monthly Notices of the Royal Astronomical Society},
  volume = {339},
  number = {3},
  pages = {623--632},
  publisher = {Oxford University Press (OUP)},
  issn = {0035-8711, 1365-2966},
  doi = {10.1046/j.1365-8711.2003.06141.x},
  urldate = {2025-07-17}
}

@article{Lyutikov1999,
  author = {Lyutikov, M.},
  title = {Beam instabilities in a magnetized pair plasma},
  journal = {Journal of Plasma Physics},
  volume = {62},
  number = {1},
  pages = {65--87},
  year = {1999},
  doi = {10.1017/S0022377899007837}
}

@article{MAHMOODFully2002,
  title = {Fully Nonlinear Dust Kinetic {{Alfv{\'e}n}} Waves},
  author = {Mahmood, M. Ansar and Mirza, Arshad M. and Sakanaka, P. H. and Murtaza, G.},
  year = 2002,
  month = sep,
  journal = {Physics of Plasmas},
  volume = {9},
  number = {9},
  pages = {3794--3801},
  issn = {1070-664X, 1089-7674},
  doi = {10.1063/1.1494983},
  urldate = {2025-10-29}
}

@article{Manchester2005,
  title = {The {{Australia Telescope National Facility Pulsar Catalogue}}},
  author = {Manchester, R. N. and Hobbs, G. B. and Teoh, A. and Hobbs, M.},
  year = 2005,
  month = apr,
  journal = {The Astronomical Journal},
  volume = {129},
  number = {4},
  pages = {1993},
  publisher = {IOP Publishing},
  issn = {1538-3881},
  doi = {10.1086/428488},
  urldate = {2026-04-06}
}

@article{Melikidze2000,
  title = {The {{Spark-associated Soliton Model}} for {{Pulsar Radio Emission}}},
  author = {Melikidze, George I. and Gil, Janusz A. and Pataraya, Avtandil D.},
  year = 2000,
  month = dec,
  journal = {The Astrophysical Journal},
  volume = {544},
  number = {2},
  pages = {1081},
  issn = {0004-637X},
  doi = {10.1086/317220},
  urldate = {2026-04-06}
}

@article{MOUSAVIDispersion2025,
  title = {Dispersion Properties of Neutron Star Magnetospheric Plasmas with Relativistic Kappa Distribution},
  author = {Mousavi, M. and Ben{\'a}{\v c}ek, J.},
  year = {2025},
  month = apr,
  journal = {Physics of Plasmas},
  volume = {32},
  number = {4},
  pages ={042111},
  publisher = {AIP Publishing},
  issn = {1070-664X},
  doi = {10.1063/5.0242852},
  urldate = {2025-07-19}
}

@article{PETRIGlobalStatic2002a,
  title = {Global Static Electrospheres of Charged Pulsars},
  author = {P{\'e}tri, J. and Heyvaerts, J. and Bonazzola, S.},
  year = {2002},
  month = mar,
  journal = {Astronomy \& Astrophysics},
  volume = {384},
  number = {2},
  pages = {414--432},
  publisher = {EDP Sciences},
  issn = {0004-6361, 1432-0746},
  doi = {10.1051/0004-6361:20020044},
  urldate = {2025-07-15}
}

@article{PHILIPPOVPulsar2022,
  title = {Pulsar {{Magnetospheres}} and {{Their Radiation}}},
  author = {Philippov, A. and Kramer, M.},
  year = 2022,
  month = aug,
  journal = {Annual Review of Astronomy and Astrophysics},
  volume = {60},
  number = {1},
  pages = {495--558},
  issn = {0066-4146, 1545-4282},
  doi = {10.1146/annurev-astro-052920-112338},
  urldate = {2025-08-24}
}

@article{PLOTNIKOVKinetic2024,
  title = {Kinetic Simulations of Electron--Positron Induced Streaming Instability in the Context of Gamma-Ray Halos around Pulsars},
  author = {Plotnikov, Illya and Van Marle, Allard Jan and Gu{\'e}pin, Claire and Marcowith, Alexandre and Martin, Pierrick},
  year = {2024},
  month = aug,
  journal = {Astronomy \& Astrophysics},
  volume = {688},
  pages = {A134},
  issn = {0004-6361, 1432-0746},
  doi = {10.1051/0004-6361/202449661},
  urldate = {2025-08-23}
}

@article{PROTHEROEGammaraysNeutrinos1998,
  title = {Gamma-Rays and Neutrinos from Very Young Supernova Remnants},
  author = {Protheroe, R. J. and Bednarek, W. and Luo, Q.},
  year = {1998},
  month = jun,
  journal = {Astroparticle Physics},
  volume = {9},
  number = {1},
  pages = {1--14},
  issn = {0927-6505},
  doi = {10.1016/S0927-6505(98)00014-0},
  urldate = {2025-07-15}
}

@article{Ruderman75,
  author  = {M.~A.~Ruderman and P.~G.~Sutherland},
  title   = {Theory of pulsars: Polar caps, sparks, and coherent microwave radiation},
  journal = {The Astrophysical Journal},
  volume  = {196},
  year    = {1975},
  pages   = {51},
}

@article{Sah2010,
  title = {Nonlinear Excitations of Kinetic {{Alfv\'en}} Waves in Electron-Positron-Ion Plasmas},
  author = {Sah, O. P.},
  year = 2010,
  month = mar,
  pages ={032306},
  journal = {Physics of Plasmas},
  volume = {17},
  number = {3},
  publisher = {AIP Publishing},
  issn = {1070-664X},
  doi = {10.1063/1.3325299},
  urldate = {2026-04-06}
}

@article{Shklovskii1970,
  author       = {Shklovskii, I. S.},
  title        = {Possible Causes of the Secular Increase in Pulsar Periods},
  journal      = {Soviet Astronomy},
  year         = {1970},
  volume       = {13},
  pages        = {562--565},
  note         = {Translated from Astronomicheskii Zhurnal, Vol. 46},
  url          = {https://ui.adsabs.harvard.edu/abs/1970SvA....13..562S/abstract}
}

@article{Shukla2004,
  title = {Low-{{Frequency Electromagnetic Waves}} in a {{Magnetized Electron-Positron-Ion Plasma}}},
  author = {Shukla, P. K. and Mendon{\c c}a, J. T. and Bingham, R.},
  year = 2004,
  month = jan,
  journal = {Physica Scripta},
  volume = {2004},
  number = {T113},
  pages = {133},
  publisher = {IOP Publishing},
  issn = {1402-4896},
  doi = {10.1238/Physica.Topical.113a00133},
  urldate = {2026-04-06}
}

@article{Singh2022,
  title = {Periodic Kinetic {A}lfv\'{e}n Waves and Overtaking between Multi-Solitons in Nonthermal Electron–Positron–Ion Plasma},
  author = {Singh, Kuldeep and Singh, Gursirat and Saini, N. S.},
  journal = {Chinese Journal of Physics},
  volume = {77},
  pages = {2060--2072},
  year = {2022},
  doi = {10.1016/j.cjph.2021.12.004}
}

@article{singhKineticAlfvenSolitary2019a,
  title = {Kinetic {{Alfv{\'e}n}} Solitary Waves in a Plasma with Two-Temperature Superthermal Electron Populations: The Case of {{Saturn}}'s Magnetosphere},
  shorttitle = {Kinetic {{Alfv{\'e}n}} Solitary Waves in a Plasma with Two-Temperature Superthermal Electron Populations},
  author = {Singh, Manpreet and Saini, N S and Kourakis, I},
  year = {2019},
  month = jul,
  journal = {Monthly Notices of the Royal Astronomical Society},
  volume = {486},
  number = {4},
  pages = {5504--5518},
  issn = {0035-8711, 1365-2966},
  doi = {10.1093/mnras/stz1221},
  urldate = {2025-06-24}
}

@article{singhKineticAlfvenicCnoidal2024,
  title = {Kinetic {{Alfv{\'e}nic}} Cnoidal Waves in {{Saturnian}} Magnetospheric Plasmas},
  author = {Singh, Manpreet and Singh, Kuldeep and Saini, N. S.},
  year = {2024},
  month = nov,
  journal = {Waves in Random and Complex Media},
  volume = {34},
  number = {6},
  pages = {5988--6001},
  issn = {1745-5030, 1745-5049},
  doi = {10.1080/17455030.2021.2015083},
  urldate = {2025-06-24}
}

@article{Singla2024,
  title = {Higher Order Corrections to Kinetic {A}lfv\'{e}n Waves in Nonthermal Plasma},
  author = {Singla, Sunidhi and Slathia, Geetika and Kaur, Rajneet and Saini, N. S.},
  journal = {IEEE Transactions on Plasma Science},
  volume = {52},
  number = {7},
  pages = {2460--2466},
  year = {2024},
  doi = {10.1109/TPS.2024.3454816}
}

@article{SIRONIPlasmoids2016,
  title = {Plasmoids in Relativistic Reconnection, from Birth to Adulthood: First They Grow, Then They Go},
  shorttitle = {Plasmoids in Relativistic Reconnection, from Birth to Adulthood},
  author = {Sironi, Lorenzo and Giannios, Dimitrios and Petropoulou, Maria},
  year = 2016,
  month = oct,
  journal = {Monthly Notices of the Royal Astronomical Society},
  volume = {462},
  number = {1},
  pages = {48--74},
  issn = {0035-8711},
  doi = {10.1093/mnras/stw1620},
  urldate = {2026-03-23}
}

@article{SPENCERDeriving2025a,
  title = {Deriving Pulsar Pair-Production Multiplicities from Pulsar Wind Nebulae Using {{H}}.{{E}}.{{S}}.{{S}}. and {{LHAASO}} Observations},
  author = {Spencer, S. T. and Mitchell, A. M. W.},
  year = {2025},
  month = feb,
  journal = {Astronomy \& Astrophysics},
  volume = {694},
  pages = {A324},
  publisher = {EDP Sciences},
  issn = {0004-6361, 1432-0746},
  doi = {10.1051/0004-6361/202451276},
  urldate = {2025-07-17}
}

@article{Stasiewicz2000,
  author  = {Stasiewicz, K. and Bellan, P. and Chaston, C. and 
             Kletzing, C. and Lysak, R. and Maggs, J. and 
             Pokhotelov, O. and Seyler, C. and Shukla, P. and 
             Stenflo, L. and Streltsov, A. and Wahlund, J.-E.},
  title   = {Small Scale {A}lfv{\'e}nic Structure in the Aurora},
  journal = {Space Science Reviews},
  year    = {2000},
  volume  = {92},
  number  = {3--4},
  pages   = {423--533},
  doi     = {10.1023/A:1005207202143}
}

@article{TIMOKHINMaximum2019,
  title = {On the {{Maximum Pair Multiplicity}} of {{Pulsar Cascades}}},
  author = {Timokhin, A. N. and Harding, A. K.},
  year = {2019},
  month = jan,
  journal = {The Astrophysical Journal},
  volume = {871},
  number = {1},
  pages = {12},
  publisher = {The American Astronomical Society},
  issn = {0004-637X},
  doi = {10.3847/1538-4357/aaf050},
  urldate = {2025-07-17}
}

@article{Urpin2011,
  author = {Urpin, {V.}},
  title = {Magnetohydrodynamic waves in the pulsar magnetosphere},
  journal = {Astronomy \& Astrophysics},
  volume = {535},
  pages = {L5},
  year = {2011},
  doi = {10.1051/0004-6361/201118047}
}

@article{Urpin2012,
  author = {Urpin, V.},
  title = {Force-free pulsar magnetosphere: instability and generation of {MHD} waves},
  journal = {Astronomy \& Astrophysics},
  volume = {541},
  pages = {A117},
  year = {2012},
  doi = {10.1051/0004-6361/201219035}
}

@article{UZDENSKYPHYSICAL2013,
  title = {{{PHYSICAL CONDITIONS IN THE RECONNECTION LAYER IN PULSAR MAGNETOSPHERES}}},
  author = {Uzdensky, Dmitri A. and Spitkovsky, Anatoly},
  year = 2014,
  month = dec,
  journal = {The Astrophysical Journal},
  volume = {780},
  number = {1},
  pages = {3},
  publisher = {The American Astronomical Society},
  issn = {0004-637X},
  doi = {10.1088/0004-637X/780/1/3},
  urldate = {2026-03-23}
}

@article{VEGARelativistic2024a,
  title = {Relativistic {{Alfv{\'e}n Turbulence}} at {{Kinetic Scales}}},
  author = {Vega, Cristian and Boldyrev, Stanislav and Roytershteyn, Vadim},
  year = {2024},
  month = apr,
  journal = {The Astrophysical Journal},
  volume = {965},
  number = {1},
  pages = {27},
  publisher = {American Astronomical Society},
  issn = {0004-637X, 1538-4357},
  doi = {10.3847/1538-4357/ad2e02},
  urldate = {2025-07-28}
}

@article{Weatherall1998,
  author = {Weatherall, J. C.},
  title = {Pulsar radio emission by conversion of plasma wave turbulence: nanosecond time structure},
  journal = {The Astrophysical Journal},
  volume = {506},
  pages = {341--346},
  year = {1998},
  doi = {10.1086/306218}
}

@article{YuLiu2021,
  title = {Solitary Dispersive {{Alfven}} Wave in a Plasma with Hot Electrons, Cold Positrons and Ions},
  author = {Yu, Y. and Liu, Y.},
  year = 2021,
  month = jul,
  journal = {Indian Journal of Physics},
  volume = {95},
  number = {7},
  pages = {1533--1543},
  issn = {0974-9845},
  doi = {10.1007/s12648-020-01798-0},
  urldate = {2026-04-06}
}

@article{Yuan2021,
  title = {Alfv\'en {{Wave Mode Conversion}} in {{Pulsar Magnetospheres}}},
  author = {Yuan, Yajie and Levin, Yuri and Bransgrove, Ashley and Philippov, Alexander},
  year = 2021,
  month = feb,
  journal = {The Astrophysical Journal},
  volume = {908},
  number = {2},
  pages = {176},
  publisher = {The American Astronomical Society},
  issn = {0004-637X},
  doi = {10.3847/1538-4357/abd405},
  urldate = {2026-04-06}
}

@book{Michel91,
  author    = {Michel, F. C.},
  title     = {Theory of Neutron Star Magnetospheres},
  publisher = {University of Chicago Press},
  year      = {1991},
}

@article{GOLDREICHPulsar1969,
  title = {Pulsar {{Electrodynamics}}},
  author = {Goldreich, P. and Julian, W. H.},
  year = {1969},
  month = aug,
  journal = {The Astrophysical Journal},
  volume = {157},
  pages = {869},
  issn = {0004-637X},
  doi = {10.1086/150119},
  urldate = {2025-08-23}
}

@article{Lyutikov2000,
  author = {Lyutikov, M.},
  title = {Excitation of {A}lfv\'{e}n Waves and Pulsar Radio Emission},
  journal = {Monthly Notices of the Royal Astronomical Society},
  volume = {315},
  pages = {31--40},
  year = {2000},
  doi = {10.1046/j.1365-8711.2000.03402.x}
}

@article{Gogoberidze2008,
  author = {Gogoberidze, G. and Machabeli, G. and Usov, V.},
  title = {On the excitation of low-frequency electromagnetic waves in pulsar magnetospheres},
  journal = {Physical Review E},
  volume = {77},
  pages = {037402},
  year = {2008},
  doi = {10.1103/PhysRevE.77.037402}
}

@book{Kadomtsev1965,
  author    = {B. B. Kadomtsev},
  title     = {Plasma Turbulence},
  year      = {1965},
  publisher = {Academic Press},
  address   = {New York},
  pages     = {82}
}

@incollection{kirk2009pulsarwinds,
  author       = {Kirk, J.G. and Lyubarsky, Y. and Petri, J.},
  title        = {The Theory of Pulsar Winds and Nebulae},
  booktitle    = {Neutron Stars and Pulsars},
  editor       = {Becker, W.},
  series       = {Astrophysics and Space Science Library},
  volume       = {357},
  pages = {421-450},
  year         = {2009},
  publisher    = {Springer},
  address      = {Berlin, Heidelberg},
  doi          = {10.1007/978-3-540-76965-1_16},
  url          = {https://doi.org/10.1007/978-3-540-76965-1_16}
}
\bibliographystyle{aasjournal}

\end{document}